SDnDTI: Self-supervised deep learning-based denoising for diffusion tensor MRI

Qiyuan Tian[1, 2]*, Ziyu Li[3], Qiuyun Fan[1, 2], Jonathan R. Polimeni[1, 2, 4], Berkin Bilgic[1, 2, 4], David H. Salat[1, 2],

Susie Y. Huang[1, 2, 4]

[1]Athinoula A. Martinos Center for Biomedical Imaging, Massachusetts General Hospital, Charlestown,

Massachusetts, United States;

[2]Department of Radiology, Harvard Medical School, Boston, Massachusetts, United States;

[3]Department of Biomedical Engineering, Tsinghua University, Beijing, P. R. China;

[4]Harvard-MIT Division of Health Sciences and Technology, Massachusetts Institute of Technology,

Cambridge, Massachusetts, United States.

*Correspondence to: Qiyuan Tian, Ph.D., Athinoula A. Martinos Center for Biomedical Imaging, 149 13th

Street, Charlestown, MA, 02129, United States. E-mail: qtian@mgh.harvard.edu.

**Abstract**

Diffusion tensor magnetic resonance imaging (DTI) is a widely adopted neuroimaging method for the in vivo mapping of brain tissue microstructure and white matter tracts. Nonetheless, the noise in the diffusion-weighted images (DWIs) decreases the accuracy and precision of DTI derived microstructural parameters and leads to prolonged acquisition time for achieving improved signal-to-noise ratio (SNR). Deep learning-based image denoising using convolutional neural networks (CNNs) has superior performance but often requires additional high-SNR data for supervising the training of CNNs, which reduces the feasibility of supervised learning-based denoising in practice. In this work, we develop a self-supervised deep learning-based method entitled "SDnDTI" for denoising DTI data, which does not require additional high-SNR data for training. Specifically, SDnDTI divides multi-directional DTI data into many subsets, each consisting of six DWI volumes along optimally chosen diffusion-encoding directions that are robust to noise for the tensor fitting, and then synthesizes DWI volumes along all acquired diffusion-encoding directions from the diffusion tensors fitted using each subset of the data as the input data of CNNs. On the other hand, SDnDTI synthesizes DWI volumes along acquired diffusion-encoding directions with higher SNR from the diffusion tensors fitted using all acquired data as the training target. SDnDTI removes noise from each subset of synthesized DWI volumes using a deep 3-dimensional CNN to match the quality of the cleaner target DWI volumes and achieves even higher SNR by averaging all subsets of denoised data. The denoising efficacy of SDnDTI is demonstrated in terms of the similarity of output images and resultant DTI metrics comparing to the ground truth generated using substantially more DWI volumes on two datasets with different spatial resolution, b-values and number of input DWI volumes provided by the Human Connectome Project (HCP) and the Lifespan HCP in Aging. The SDnDTI results preserve image sharpness and textural details and substantially improve upon those from the raw data. The results of SDnDTI are comparable to those from supervised learning-based denoising and outperform those from state-of-the-art conventional denoising algorithms including BM4D, AONLM and MPPCA. By leveraging domain knowledge of diffusion MRI physics, SDnDTI makes it easier to use CNN-based denoising methods in practice and has the potential to benefit a wider range of research and clinical applications that require accelerated DTI acquisition and high-quality DTI data for mapping of tissue microstructure, fiber tracts and structural connectivity in the living human brain.

**Keywords**: convolutional neural network, supervised learning, residual learning, image synthesis, diffusion tensor transformation, normal aging.

**Introduction**

Diffusion tensor magnetic resonance imaging (DTI)[1-5] is a widely adopted neuroimaging method that noninvasively maps the brain tissue properties and white matter tracts in the in vivo human brain. DTI mathematically models the multi-directional diffusion rates of water molecules measured by diffusion MRI as a tensor (i.e., a 3×3 symmetric matrix) and infers tissue microstructure based on the fact that diffusion rates in white matter are generally anisotropic due to the restriction of axonal membranes, myelin sheath and other microscopic barriers. For example, the primary eigenvector of the tensor with the largest diffusion rate often relates to the local axonal orientation, since the random motion of water molecules is less restricted along the length of fibers. On the other hand, the mean diffusion rate indicates the overall restriction of the tissue microenvironment.

DTI has a wide range of applications in research and clinical studies. Since the metrics from DTI have high sensitivity and specificity to brain tissue microstructure, DTI has proven to be a valuable tool for characterizing and monitoring microstructural changes related to development[6,7], normal aging[6,8], neurodegeneration[9], plasticity[10] and a number of neurological[11,12] and psychiatric[13,14] disorders. In clinical practice, DTI-based tractography (i.e., white matter fiber tracking) is routinely used for presurgical planning for brain tumor resection and epilepsy surgery, as well as in identifying the optimal target location for functional neurosurgery using deep brain stimulation and MRI guided focused ultrasound[15,16]. In addition to investigating white matter, high-resolution DTI also provides a valuable tool for mapping human cerebral cortical gray matter anisotropy and microstructure[17-21]. For these reasons, DTI is an essential imaging modality adopted in many large-scale neuroimaging studies such as the Alzheimer's Disease Neuroimaging Initiative[22,23], the Parkinson Progression Marker Initiative[24], and the UK Biobank Imaging Study[25].

Nonetheless, the relatively long acquisition time poses a critical barrier to performing high-quality DTI in routine clinical practice and large-scale research studies. Since diffusion-weighted MRI creates image contrast by attenuating MR signals based on how easily water molecules can diffuse in a local brain region, diffusion-weighted images (DWIs) are by nature noisy, especially in acquisitions using strong diffusion encoding (i.e., high b-values) or high spatial resolution. Consequently, DTI often needs several times more measurements (e.g., 20 DWI volumes, 5–10 seconds per volume) than the theoretical minimum of 6 DWI volumes to accurately, precisely and robustly derive the six unique elements of the tensor, leading to lengthy scans. For DTI performed at high (1.25 to 1 mm isotropic) or ultra-high (sub-millimeter isotropic) spatial resolution, the required number of measurements can be far more than 20. For example, two repetitions of DWI volumes sampled along 256 uniformly distributed directions at 1-mm isotropic spatial resolution were acquired in 65-minute DTI scans for mapping the primary fiber orientation in the cerebral cortex, which provides a myeloarchitecture-based contrast mechanism for parcellating cortical regions in vivo[17].

Image denoising provides a feasible alternative strategy to improve the quality of DTI from a shorter scan and make it equivalent to that from a longer scan. Image denoising aims to recover a clean image with high signal-to-noise ratio (SNR) from noise-degraded observations, which is a highly ill-posed inverse problem. In the computer vision field, numerous denoising algorithms have been proposed to remove noise from natural and biomedical images, such as total variation denoising[26], anisotropic diffusion filtering[27,28], bilateral filtering[29], non-local means (NLM) filtering[30], block-matching and 3-dimentional filtering (BM3D)[31] and K-singular value decomposition (K-SVD) denoising[32] and their 3-dimensional extensions for volumetric data[33-37]. Many of these algorithms are able to deal with spatially varying non-Gaussian noise and therefore can be readily applied to denoise diffusion MRI data. MRI reconstruction methods that often regularize the image formation process using prior knowledge, such as sparseness[38-42] and low rank[43-48], can also achieve denoising effects. For example, low rank constraint has been employed in k-space[49,50] and image space[43] for multi-shot diffusion MRI reconstruction to provide improved SNR. Many image restoration methods are also designed to exploit the additional redundant information originating from multiple diffusion encodings along various directions for increased denoising efficacy. A representative example is the widely adopted Marchenko–Pastur principal component analysis (MPPCA) algorithm[51,52], which isolates and suppresses the noise-only component of the spatial-diffusion signals in the eigenspectrum domain. Many variants of this algorithm based on similar principles have been also proposed[37,53,54]. Multiple DWIs have also been jointly reconstructed to exploit their inter-image correlation for enhanced SNR[44,55,56]. Another category of methods explicitly imposes a model of the signals in diffusion space to remove noise[57,58].

Emerging deep learning technologies, particularly convolutional neural networks (CNNs), offer another powerful tool set for image denoising. With supervision, CNNs can automatically learn to fully utilize the redundancy embedded in the data and effectively restore noise-free images from their noisy observations. It has been shown that the CNN for denoising (i.e., DnCNN[59]) with a simple network architecture outperforms the state-of-the-art BM3D denoising method. Therefore, CNN-based denoising has been widely adopted for fluorescence microscopy[60,61], optical coherence tomography[62], x-ray imaging[63], x-ray computed tomography[64], PET[65-69] and MRI[64,70-77]. For diffusion MRI, many studies have proven the superiority of CNNs in estimating high-quality scalar diffusion metrics from DTI[78-83] and more advanced diffusion models[79,82-84] as well as voxel-wise axonal orientations[85] from a small amount of input data for faster imaging. In parallel to these studies, the DeepDTI[76] method leverages CNNs to denoise six DWI volumes sampled along optimally selected diffusion-encoding directions with the target clean images synthesized from tensors fitted using more data, achieving approximately four-fold acceleration over non-denoised images.

Despite their superior performance, most deep learning-based denoising methods require additional high-SNR data for the supervised training of CNNs and are therefore more difficult to use in practice comparing to conventional algorithms. On the one hand, the performance of a pre-trained CNN might be compromised when it is directly applied to a new dataset acquired with different hardware systems and sequences, which exhibit different image contrast, spatial resolution, SNR level, and so on. On the other hand, to create a custom CNN optimized for the data in a new application requires additional high-SNR image data as the training target from numerous subjects, even for fine-tuning parameters of pre-trained CNNs. These high-quality training targets are usually obtained from a large number of measurements from longer scans, which might be challenging to acquire on children, older subjects, and certain populations of patients, and are also unavailable for legacy data.

To address this challenge, we present a self-supervised deep learning framework for denoising DTI data entitled "SDnDTI" that does not require external high-SNR data for training. SDnDTI employs a simple concept of "first denoising then averaging". In the simple case for interspersed $b = 0$ image volumes of a DTI dataset, SDnDTI denoises each single $b = 0$ image volume using the CNN with the averaged $b = 0$ image volume as the training target and then averages all denoised $b = 0$ image volumes. Since each denoised $b = 0$ image volume has equivalent SNR comparing to the averaged $b = 0$ image volume due to the superior denoising performance of CNNs, the average of all denoised $b = 0$ images recovers higher SNR than the averaged $b = 0$ image and thus achieves denoising effects. Since DWI volumes in a DTI dataset are sampled along uniformly distributed directions thus exhibiting different image contrast, SDnDTI leverages the diffusion tensor model to implement the "first denoising then averaging" concept for denoising DWI volumes. In order to obtain multiple repetitions of DWI volumes with identical image contrast but independent noise observations, SDnDTI divides all DWI volumes into several subsets, each with six DWI volumes along optimally chosen diffusion-encoding directions that minimize noise amplification for the tensor fitting, and then synthesizes DWI volumes along all acquired diffusion-encoding directions from the diffusion tensors fitted using each subset of the data (along with the averaged $b = 0$ image volume) as the input data of CNNs. On the one hand, SDnDTI synthesizes DWI volumes along acquired diffusion-encoding directions with higher SNR from the diffusion tensors fitted using all acquired data as the training target of CNNs. SDnDTI removes noise from each subset of synthesized DWI volumes using the CNN to match the quality of the cleaner target DWI volumes and achieves higher SNR by averaging all subsets of denoised data.

In this work, we lay out the framework for SDnDTI and then systematically quantify the similarity of denoised images and resultant DTI metrics from SDnDTI compared to the ground truth generated from substantially more DWI volumes using two separate datasets acquired with different spatial resolutions and b-values, namely, those from the Human Connectome Project (HCP) and Lifespan HCP in Aging. We show that SDnDTI results substantially improve upon those from the raw data, outperform those from state-of-the-art denoising algorithms, including BM4D[31,36], adaptive optimized NLM (AONLM)[35] and MPPCA, and are comparable to those from supervised deep learning-based denoising using external ground-truth data as the training target. Because of the superior performance and reduced requirement for training data, we anticipate easier deployment and wider use of SDnDTI in practice that might benefit a broader range of clinical and neuroscientific research.

**METHODS**

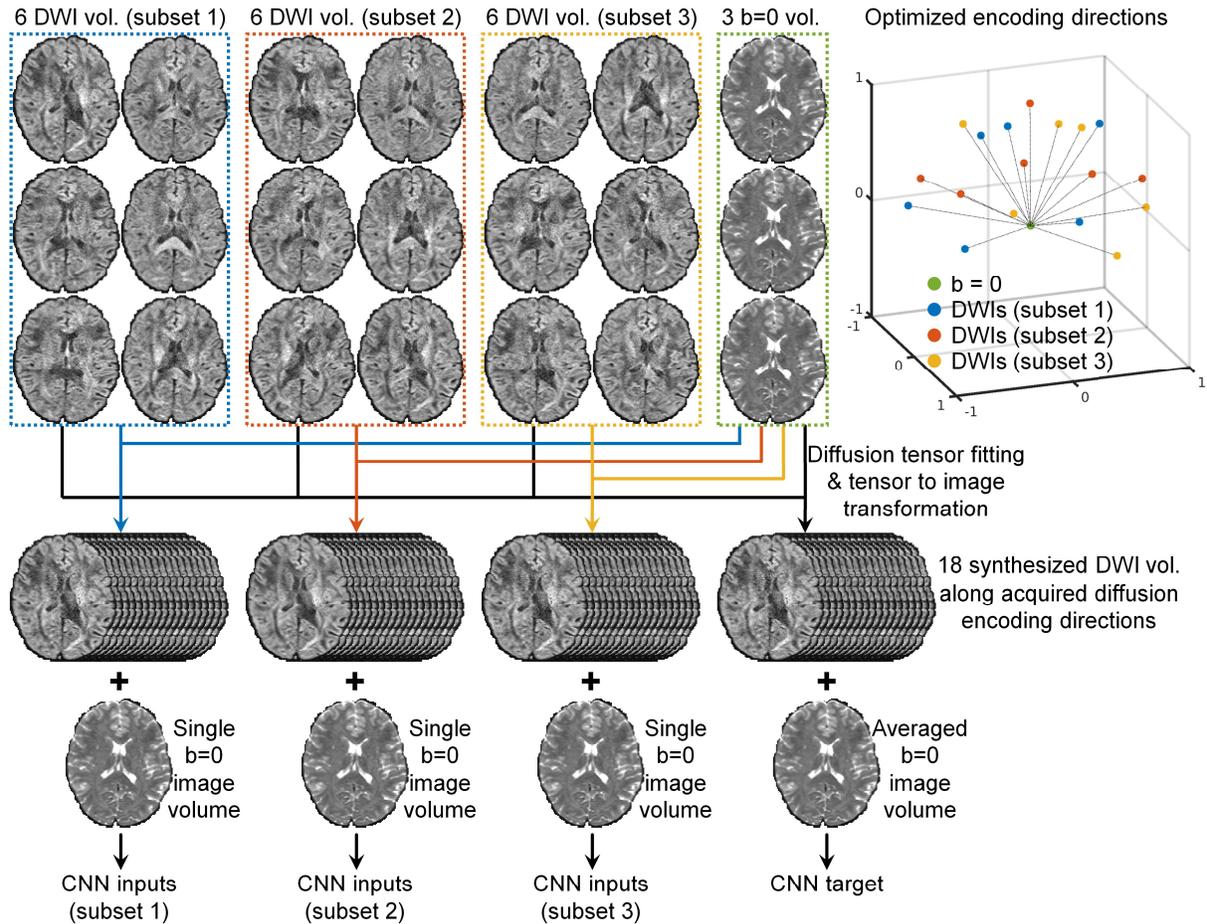

**Figure 1. SDnDTI pipeline.** The SDnDTI pipeline for a DTI acquisition consisting of three $b = 0$ image volumes and 18 diffusion-weighted image volumes is demonstrated.

SDnDTI pipeline

The SDnDTI pipeline for a DTI acquisition consisting of three b = 0 image volumes and 18 DWI volumes is demonstrated in Figure 1 and could be extended to any number of DWI volumes. The diffusion-encoding directions of the 18 DWI volumes need to be optimized such that they can be divided into 3 subsets of 6 directions which minimize the noise amplification during the diffusion tensor fitting process, and the 18 directions are also uniformly distributed on a sphere to ensure uniform angular coverage.

For each subset of 6 DWI volumes, tensor fitting is performed along with the averaged b = 0 image volume to estimate low-quality diffusion tensors, which are then used to synthesize DWI volumes sampled along the 18 acquired encoding directions. The synthesis of these DWI volumes serves to transform DWI volumes sampled along different encoding directions to the same directions while maintaining full angular coverage. A single b = 0 image volume and 18 synthesized DWI volumes serve as the inputs to the CNN. Specifically, for each voxel, the diffusion tensor $\mathbf{D_6} = \begin{bmatrix} D_{xx}\ D_{yy}\ D_{zz}\ D_{xy}\ D_{xz}\ D_{yz} \end{bmatrix}^T$ consisting of 6 unique elements is estimated using linear squares fit as:

$$\mathbf{D_6} = \mathbf{A_6}^{-1}\mathbf{C_6}, \quad\quad\quad (1)$$

where $\mathbf{C_6} = \begin{bmatrix} c_1\ c_2\ c_3\ c_4\ c_5\ c_6 \end{bmatrix}^T$ represents the apparent diffusion coefficients (ADCs) along the 6 diffusion-encoding directions, with $c_i = -\ln\left(S_i/S_0\right)/b_i$ ($i$ = 1, 2, 3, 4, 5, 6), $S_0$ as the non-diffusion-weighted signal intensity from the average of the three b = 0 image volumes, $S_i$ as the diffusion-weighted signal intensity and $b_i$ as the b-value, and $\mathbf{A_6} = \begin{bmatrix} \boldsymbol{\alpha_1}\ \boldsymbol{\alpha_2}\ \boldsymbol{\alpha_3}\ \boldsymbol{\alpha_4}\ \boldsymbol{\alpha_5}\ \boldsymbol{\alpha_6} \end{bmatrix}^T$ represents the diffusion tensor transformation matrix, with $\boldsymbol{\alpha}_i^T = \begin{bmatrix} g_{ix}^2\ g_{iy}^2\ g_{iz}^2\ 2g_{ix}g_{iy}\ 2g_{ix}g_{iz}\ 2g_{iy}g_{iz} \end{bmatrix}$ ($i$ = 1, 2, 3, 4, 5, 6) solely depending on the diffusion-encoding directions $(g_{ix}, g_{iy}, g_{iz})^T$. In order to avoid large noise amplification when solving the tensor, the six diffusion-encoding directions need to be selected to minimize the condition number of $\mathbf{A_6}$[86]. The image intensities of each voxel in the synthesized DWI volumes along the 18 acquired diffusion-encoding directions are calculated as:

$$\mathbf{S_{18}} = S_0 e^{-diag(\boldsymbol{b})\cdot\mathbf{A_{18}}\mathbf{D_6}} = S_0 e^{-diag(\boldsymbol{b})\cdot\mathbf{A_{18}}\mathbf{A_6^{-1}}\mathbf{C_6}} \quad\quad\quad (2)$$

, where $\mathbf{A_{18}}$ is the diffusion tensor transformation matrix associated with the 18 acquired directions, and *diag* stands for the diagonalization operation of a vector of the b-values $\boldsymbol{b}$ of the 18 DWI volumes. The calculated image intensities of $\mathbf{S_{18}}$ along the directions of the 6 DWI volumes used for the tensor fitting are identical to the raw acquired image intensities since the diffusion tensor transformation is well conditioned and fully invertible (except for very few outlier voxels where some DWI intensities because of noise happen to be higher than the b = 0 image intensity such as on the edge of the brain where the b = 0 image intensities are very low). Only the rest 12 computed image intensities of $\mathbf{S_{18}}$ are actually synthesized, which are different from the raw acquired image intensities and also exhibit different noise characteristics.

The tensor fitting is also performed on three b = 0 image volumes and all 18 DWI volumes to estimate a diffusion tensor with higher quality, which is then used to generate synthesized DWI volumes with higher SNR along the 18 acquired diffusion-encoding directions in a similar way as described above. The averaged b = 0 image volume and the 18 synthesized DWI volumes with higher SNR serve as the target of the CNN.

For each subject, three pairs of inputs and targets consisting of one b = 0 image volume and 18 DWI volumes are created. The b = 0 image volume and DWI volumes are jointly denoised to enhance data redundancy for boosted CNN performance. Any CNN for denoising can be trained in a supervised fashion to improve the SNR of each subset of input image volumes. The training can be performed using data from many subjects in a project which need to be denoised or from a single subject in a subject-specific fashion. Generally speaking, denoising the data of numerous subjects jointly is beneficial because the increased amount of training data can be used to train deeper CNNs with more parameters for boosted denoising performance. The denoised images of all three subsets from the CNN are averaged to obtain the final denoised results, which are then used for tensor fitting to derive scalar and orientation DTI metrics.

A simple approach can be used to select the optimal 18 diffusion-encoding directions. First, the six optimized diffusion-encoding directions from the DSM scheme[86] that minimize the condition number of the diffusion transformation matrix to 1.3228 while being as uniform as possible are chosen. Second, three sets of the six optimized directions are randomly rotated for many times and the 18 most uniform directions, i.e., with the minimal electrostatic potential energy[87], are chosen.

Modified U-Net

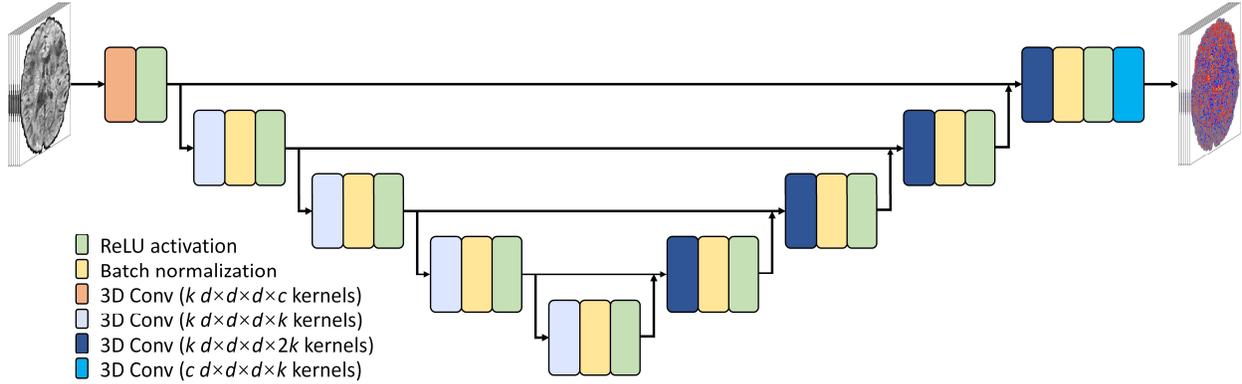

**Figure 2. Modified U-Net (MU-Net) architecture.** MU-Net is modified from U-Net by removing all max pooling and up-sampling layers and keeping the number of kernels constant across all layers. The input is $c$ noisy image volumes (one $b = 0$ image and $c$ - 1 diffusion-weighted image volumes). The output is $c$ residual images volumes between the input noisy image volumes and high-quality target image volumes. Network parameters of $k = 192$ and $d$ = 3 were adopted in this study.

A modified 10-layer 3-dimensional (3D) U-Net[88] (MU-Net) was used to learn the mapping from the input noisy image volumes to the residuals between the input and output image volumes with higher SNR (i.e., residual learning), which were then added to the input image volumes to generate resultant denoised image volumes. Specifically, all max pooling and up-sampling layers of U-Net were removed to preserve the native spatial resolution at each layer, and the number of kernels at each layer was kept constant ($k = 192$). Compared to 2D convolution, 3D convolution ($d{\times}d{\times}d = 3{\times}3{\times}3$ kernel size, $1{\times}1{\times}1$ stride) increases the data redundancy from an additional spatial dimension for improved image synthesis performance and avoids boundary artifacts between 2D image slices. Essentially, MU-Net is composed of a sequence of paired convolution, batch normalization and rectified linear unit (ReLU) activation layers with several short paths from early layers to later layers. These skip connections serve to alleviate the vanishing-gradient problem and strengthen feature propagation. MU-Net represents an intermediate network between a plain network (e.g., VDSR[89] and DnCNN[59]) without any short paths and a densely connected network (e.g., DenseNet[90]) that comprehensively connects each layer to every other layer.

Human Connectome Project data

Pre-processed diffusion MRI data of 20 unrelated young healthy subjects from the HCP WU-Minn-Ox Consortium (https://www.humanconnectome.org) were used for this study. The acquisition methods and parameter values have been described in detail previously[91-93], and those relevant to this study are briefly described below. Diffusion MRI data were acquired in the whole brain at 1.25 mm isotropic resolution using a two-dimensional diffusion-weighted pulsed-gradient spin-echo echo-planar imaging (DW-PGSE-EPI) sequence, with three b-values (1, 2, 3 ms/$\mu$m$^2$) and two phase-encoding directions (left–right and right–left). For each b-value, 90 diffusion-encoding directions uniformly distributed on a sphere were acquired[94]. The diffusion data were corrected for gradient nonlinearity, eddy current and susceptibility induced distortions and co-registered using the FMRIB Software Library (FSL) software[95-98] (https://fsl.fmrib.ox.ac.uk). The image volumes acquired with the left–right and right–left phase-encoding directions were combined into a single image volume by FSL's "eddy" function. Only the 18 interspersed combined $b = 0$ image volumes and 90 combined DWI volumes at b = 1 ms/$\mu$m$^2$ were used in this study. In addition, the volumetric segmentation results from the FreeSurfer[99,100] (https://surfer.nmr.mgh.harvard.edu) reconstruction on the T$_1$-weighted data were also used in this study to derive brain tissue masks for results evaluation.

Human Connectome Project in Aging data

The diffusion and T$_1$-weighted MRI data of 20 unrelated healthy adults (ages 36-93, mean age 65.8±19.1, 10 female) from the Lifespan HCP in Aging (HCP-A) study[101,102] were used for this study. The data were acquired as part of the HCP-A at the Massachusetts General Hospital Martinos Center for Biomedical Imaging with approval from the institutional review board and written informed consent from all participants. The subjects were randomly selected with uniformly distributed ages. The data acquisition was performed using a 3-T MRI scanner (MAGNETOM Prisma; Siemens Healthcare, Erlangen, Germany) equipped with the Siemens 32-channel Prisma head coil for signal reception.

Whole-brain diffusion data were acquired at 1.5 mm isotropic resolution using a two-dimensional DW-PGSE-EPI sequence with the following parameters: repetition time = 3,230 ms, echo time = 89.2 ms, contiguous axial slices, simultaneous multi-slice factor = 4, without in-plane acceleration, with two b-values (1.5, 3 ms/μm$^2$) and two phase-encoding directions (anterior–posterior and posterior–anterior). For b=1.5 ms/μm$^2$ and b=3 ms/μm$^2$, 93 and 92 diffusion-encoding directions uniformly distributed on a sphere were acquired. Only the 28 interspersed $b = 0$ image volumes and 186 DWI volumes at b = 1.5 ms/μm$^2$ were used in this study.

Whole-brain T$_1$-weighted data were acquired at 0.8-mm isotropic resolution using a 3-dimensional sagittal multi-echo magnetization-prepared rapid acquisition with gradient echo (ME-MPRAGE) sequence[103] with the following parameters: repetition time = 2,500 ms, echo time = 1.8/3.6/5.4/7.2 ms, inversion time = 1000 ms, flip angle = 8°, partial Fourier factor = 6/8, generalized autocalibrating partial parallel acquisition (GRAPPA) factor = 2.

Human Connectome Project in Aging data Processing

The HCP-A diffusion data were corrected for eddy current and susceptibility induced distortions and co-registered using the "topup" and "eddy" functions from the FSL software. In order to compare to results from the MPPCA denoising method[51,52] which needs to be applied to the unprocessed raw data from the scanner, the resultant warp field maps for correcting and aligning each image volume from the "eddy" function were saved out with the "--dfields" option. Each warp field map was applied to individual image volume from the MPPCA-denoised raw data using FSL's "applywarp" function with the "spline" interpolation to obtain the distortion-free and co-registered MPPCA-denoised data. Each warp field map was also applied to individual image volume from the raw diffusion data using FSL's "applywarp" function with the "spline" interpolation to obtain the distortion-free and co-registered raw data, which were used for the subsequent image processing and denoising. The image volumes acquired with the anterior–posterior and posterior–anterior phase-encoding directions were averaged and combined into a single image volume to account for the signal loss in the brain regions with large susceptibility induced distortions due to the absence of in-plane acceleration, resulting in 14 combined $b$ = 0 image volumes and 93 combined DWI volumes at b = 1.5 ms/μm². The corrected data directly from the "eddy" function were not used, because the "eddy" function internally replaces the detected outlier image slices with its own estimation, which introduces a confounding factor for the comparison of MPPCA and other denoising methods.

FreeSurfer reconstruction was performed on the $T_1$-weighted data using the "recon-all" function.

Image processing

For each subject, the diffusion data were corrected for spatially varying intensity biases using the averaged b = 0 image volume with the unified segmentation routine implementation in the Statistical Parametric Mapping software (SPM, https://www.fil.ion.ucl.ac.uk/spm) with a full-width at half-maximum of 60 mm and a sampling distance of 2 mm.

The volumetric brain segmentation results (i.e., aparc+aseg.mgz) from FreeSurfer reconstruction on the $T_1$-weighted data were re-sampled to the diffusion image space with an affine transformation using nearest neighbor interpolation. For the HCP data, the diffusion data and the $T_1$-weighted were co-registered already and therefore the affine transformation was simply an identity matrix. For each HCP-A subject, the affine transformation was derived using the averaged b = 0 image volume with FreeSurfer's "bbregister" function[104]. Binary masks of brain tissue that excluded the cerebrospinal fluid (CSF) were obtained using FreeSurfer's "mri_binarize" function with "--gm" and "--all-wm" options which were used for evaluating results.

Input data selection

Three b = 0 image volumes and 18 DWI volumes of each subject from HCP and two b = 0 image volumes and 12 DWI volumes of each subject from HCP-A were used to demonstrate the denoising efficacy of SDnDTI on input data with different spatial resolution, b-values and number of input DWI volumes as well as compare to other denoising methods. Because the diffusion-encoding directions of the pre-acquired HCP and HCP-A data were fixed, it was impossible to obtain the optimal 6 diffusion-encoding directions from the DSM scheme or their rotations that minimized the condition number of the diffusion transformation matrix to 1.3228. Therefore, all possible sets of six diffusion-encoding directions that associate with a diffusion tensor transformation matrix with a condition number lower than 1.6 were used (31 sets for HCP data and 46 sets for HCP-A data). These sets of diffusion-encoding directions were selected by randomly rotating the 6 optimal directions from the DSM scheme to six new directions and then keeping the set of the six nearest directions if their associated condition number was lower than 1.6. Then three out of 31 sets of six directions for the HCP data and two out of 46 sets of six directions for the HCP-A data were randomly picked many times, and the 18 or 12 selected directions with the lowest electrostatic potential energy[105] were chosen, which were uniformly distributed on a sphere (Figure 1, Supplementary Figure 1).

<u>SDnDTI denoising implementation</u>

The MU-Net of SDnDTI was implemented using the Keras application programming interface (https://keras.io) with a Tensorflow backend (https://www.tensorflow.org). The mean absolute error (MAE, i.e., L1 loss) was used to optimize the CNN parameters using the Adam optimizer[106] with default parameters (except for the learning rate). The learning rate was empirically set to 0.0001. Only the MAE within the brain mask was used.

To account for subject-to-subject variations in image intensity, the intensities of the input and target images of SDnDTI were standardized by subtracting the mean image intensity and dividing by the standard deviation of image intensities across all voxels within the brain mask from the input images. Input and target images were brain masked. The training data were flipped along the anatomical left-right direction to augment the training data. All 3D convolutional kernels were randomly initialized with a "He" initialization[107]. Blocks of 64×64×64 voxel size were used for training (8 blocks from each subject) due to the limited memory of graphics processing unit (GPU).

The training and validation were performed on 20 subjects using a Tesla V100 GPU with 16 GB memory (NVIDIA, Santa Clara, CA). For each epoch, 80% randomly selected blocks were used for training and the remaining 20% were used for validation. The batch size was set to one, the largest size that can be accommodate by the GPU. The training and validation were performed for 40 epochs for the HCP data (~60 minutes per epoch) and 60 epochs for the HCP-A data (~30 minutes per epoch). During the training, the training error kept decreasing while the validation error decreased first, reached the minimum and then started increasing when the CNN parameters started to be over fitted. The MU-Net parameters with the minimum validation errors were used (from the 34th epoch for the HCP data and the 19th epoch for the HCP-A data).

The learned network parameters were applied to the whole brain volume of each subject. The standardized image intensities were transformed back to the normal range by multiplying with the standard deviation of image intensities across all voxels within the brain mask from the input images and then add the mean image intensity of brain voxels.

The network implementation and training parameters (e.g., learning rate, the way to select the suitable number of training epoch based on tracking the validation error) were kept the same in the following sections if not explicitly specified.

Effects of the amount of training data

Experiments were performed on the HCP data to assess the effects of the amount of training data on the SDnDTI performance. For the HCP data, the MU-Net of SDnDTI was trained on data from subsets of 10 subjects, 5 subjects, or 1 subject out of the 20 subjects in the similar way as described above, resulting in 2, 4, and 20 networks with optimized parameters respectively. Each optimized MU-Net was then applied to the data of subjects used for its training to denoise images.

For the training using 10 subjects, the training and validation were performed for 80 epochs (~60 minutes per epoch) and the networks from the 50th and 57th epoch were used for denoising the data from the first 10 subjects and the last 10 subjects, respectively. For the training using 5 subjects, the training and validation were performed for 80 epochs (~15 minutes per epoch) and the networks from the 52nd, 47th, 52nd and 43rd epoch were used for denoising the data from each quarter of the 20 subjects, respectively. For the training using 1 subject, the training and validation were performed for 100 epochs (~3 minutes per epoch) and the networks from the 69th, 84th, 41st, 98th, 57th, 88th, 81st, 89th, 86th, 98th, 89th, 79th, 98th, 96th, 78th, 77th, 79th, 72nd, 95th, 79th epoch were used for denoising the data from each of the 20 subjects.

Network generalization and fine tuning

In order to evaluate the generalization of the MU-Net of SDnDTI, two b = 0 image volumes and 12 DWI volumes of each of the 20 subjects from HCP (1.25 mm isotropic resolution, b = 1 ms/$\mu$m$^2$) were selected and used to train a MU-Net in the same way as described above. In order to be directly applied to the HCP-A data, the input and target DWIs were synthesized along the encoding directions from the HCP-A input data from the tensor fitted using each subset of the HCP data and all selected data, respectively. The training and validation were performed for 40 epochs (~40 minutes per epoch) and the network from the 30th epoch was directly applied to denoise the data (1.5 mm isotropic resolution, b = 1.5 ms/$\mu$m$^2$) from each of the HCP-A subjects.

The data from the HCP-A were used for fine-tuning parameters of the pre-trained MU-Net using the HCP data. Specifically, for each of the 20 subjects from the HCP-A, another MU-Net, initialized with parameters of the MU-Net learned from the HCP data, were further fine-tuned on the data of this subject. The training and validation were performed for 40 epochs (~1 minute per epoch) and the networks from the 23rd, 12nd,

32nd, 18th, 23rd, 29th, 23rd, 23rd, 34th, 20th, 30th, 14th, 17th, 29th, 27th, 30th, 33rd, 19th, 21st, 8th epoch were used for denoising the data from each of the 20 HCP-A subjects.

For comparison, an MU-Net was also trained from randomly initialized parameters on the data from each of the 20 HCP-A subjects. The training and validation were performed for 80 epochs (~1 minute per epoch) and the networks from the 40th, 30th, 41st, 36th, 48th, 37th, 38th, 46th, 65th, 44th, 30th, 33rd, 40th, 35th, 37th, 42nd, 43rd, 36th, 41st, 25th were used to denoise the data from each of the 20 HCP-A subjects. Comparing to the fine tuning, training from random initialization required more epochs to converge.

<u>Denoising using other methods</u>

For comparison, diffusion data were also denoised using three state-of-the-art traditional denoising methods, including BM4D, AONLM and MPPCA, as well as using supervised learning with external high-SNR data. BM4D and AONLM were applied to pre-processed HCP and HCP-A diffusion data. The MPPCA was only applied to the unprocessed HCP-A data from the scanner. The MPPCA-denoised data were then corrected for distortions and co-registered as described above.

The BM4D denoising, an extension of the well-known BM3D algorithm for volumetric data, was set to estimate the unknown noise standard deviation of the Rician noise and perform collaborative Wiener filtering with "modified profile" option and default parameters using the publicly available MATLAB-based software ([https://www.cs.tut.fi/~foi/GCF-BM3D](https://www.cs.tut.fi/~foi/GCF-BM3D)). The AONLM was performed assuming Rician noise with $3 \times 3 \times 3$ block and $7 \times 7 \times 7$ search volume[34,35] using the publicly available MATLAB-based software ([https://sites.google.com/site/pierrickcoupe/softwares/denoising-for-medical-imaging/mri-denoising/mri-denoising-software](https://sites.google.com/site/pierrickcoupe/softwares/denoising-for-medical-imaging/mri-denoising/mri-denoising-software)). The MPPCA denoising was performed with $5 \times 5 \times 5$ kernel and "full" sampling using the publicly available MATLAB-based software ([https://github.com/NYU-DiffusionMRI/mppca_denoise](https://github.com/NYU-DiffusionMRI/mppca_denoise)).

Supervised learning-based denoising using MU-Net with external high-SNR data as the training target was also performed for comparison. In this case, the input of the MU-Net is the raw acquired three b = 0 image volumes and 18 DWI volumes or two b = 0 image volumes and 12 DWI volumes with low SNR for the HCP and HCP-A data, respectively. The output of the MU-Net is three b = 0 image volumes and 18 DWI volumes or two b = 0 image volumes and 12 DWI volumes with high SNR for the HCP and HCP-A data, respectively. For each subject, the high-SNR b = 0 image volume was computed by averaging all available b = 0 image volumes (i.e., 18 or 14 volumes for each HCP or HCP-A subject, respectively). The high-SNR DWI volumes were synthesized from the diffusion tensor generated using all diffusion data (i.e., 18 b = 0 image volumes and 90 DWI volume for each HCP subject, 14 b = 0 image volumes and 93 DWI volumes for each HCP-A subject) as described in Equation 2. For both the HCP and HCP-A data, the training and validation were performed on the data from 20 subjects for 40 epochs (~21 minutes per epoch for the HCP data and ~14 minutes per epoch for the HCP-A data). The MU-Net from the 31st epoch and the 32nd epoch were used for denoising the HCP and HCP-A data, respectively.

Quantitative comparison

Resultant denoised images and the DTI metrics including the primary eigenvector (V1), fractional anisotropy (FA), mean diffusivity (MD), axial diffusivity (AD), and radial diffusivity (RD) were compared to the ground truth for evaluating the denoising performance. As for the supervised learning-based denoising, the ground-truth b = 0 image volumes were computed by averaging all available b = 0 image volumes for each subject and the ground-truth DWI volumes were synthesized from the diffusion tensor generated using all available diffusion data. The diffusion tensor fitting was performed using ordinary linear squares fitting using FSL's "dtifit" function to derive the diffusion tensor, V1, FA, MD, AD and RD.

The MAE, Peak SNR (PSNR) and structural similarity index (SSIM)[108] were used to quantify the similarity between the raw input images and denoised images using different methods compared to the ground-truth images. For these calculations, the image intensities of BM4D-, AONLM- and MPPCA-denoised image volumes were first standardized in the same way as for preparing the input and output data for SDnDTI. The standardized image intensities for denoised image volumes from all methods, within the range [-3, 3], were transformed to the range of [0, 1] by adding 3 and dividing by 6. PSNR was computed using MATLAB's "psnr" function, with larger value indicating lower mean squared error. SSIM was computed using MATLAB's "ssim" function, with larger value indicating high perceptual similarity. The group mean and group standard deviation of the MAE, PSNR and SSIM across the 20 subjects from HCP and HCP-A were calculated, respectively.

The mean absolute differences (MAD) of V1, FA, MD, AD, RD compared to the ground-truth DTI metrics within the brain (excluding the cerebrospinal fluid) were used to quantify the accuracy of the DTI results. The difference of V1 was computed as the angle (between 0 and 90°) between the two primary eigenvectors for comparison. The group mean and group standard deviation of the mean values of the MAD for different metrics across the 20 subjects from HCP and HCP-A were calculated, respectively.

# Results

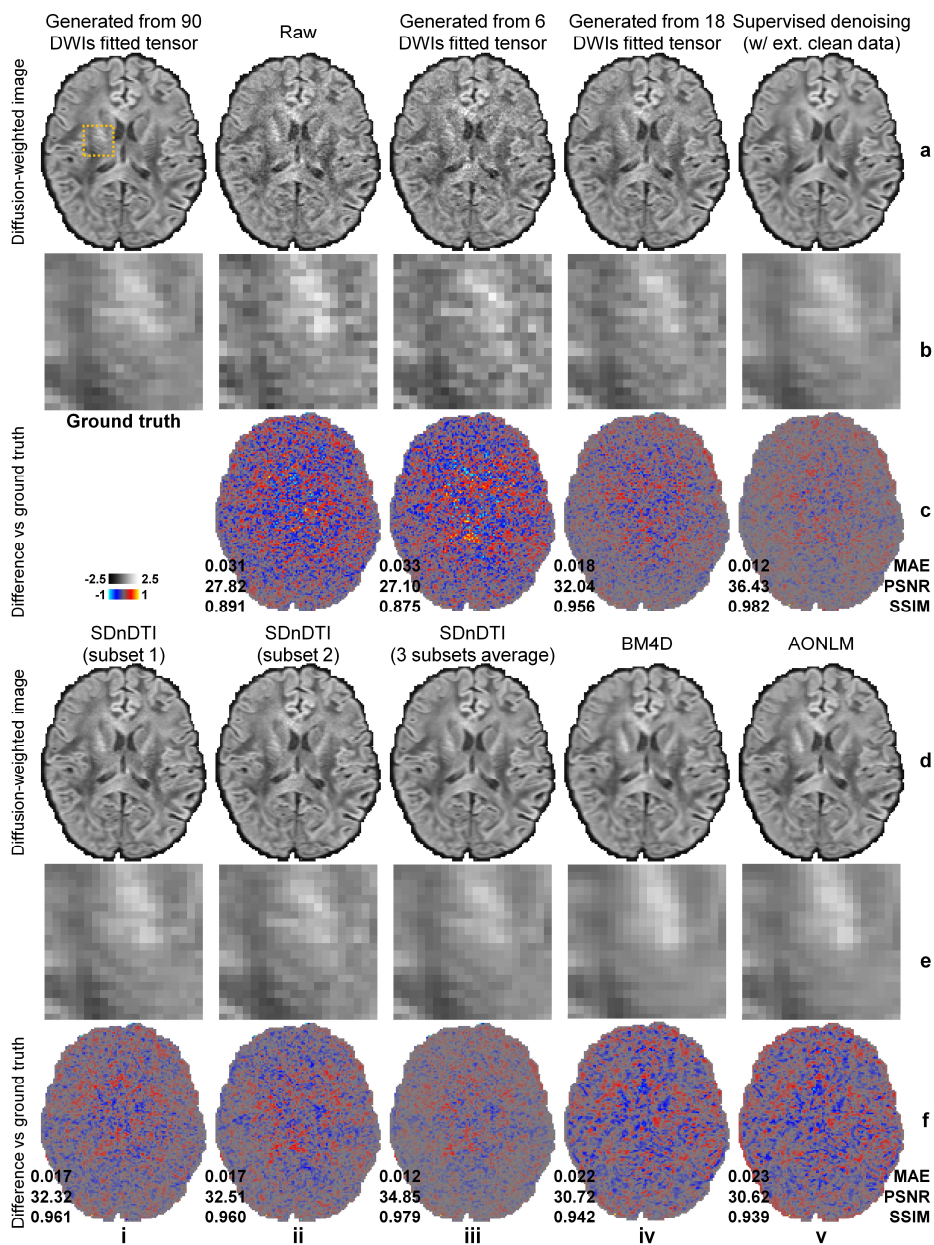

**Figure 3. Denoised images.** Diffusion-weighted images (DWIs) sampled approximately along the superior-inferior direction (i.e., [-0.18, 0.26, -0.95]) of a representative HCP subject from the ground-truth data (i.e., synthesized from the diffusion tensor fitted using all 18 b = 0 images and 90 DWIs) (a, i), subset 1 of SDnDTI input data (i.e., raw acquired image) (a, ii), subset 2 of SDnDTI input data (i.e., synthesized image using the diffusion tensor fitted using three b = 0 images and six DWIs) (a, iii), synthesized data from the diffusion tensor fitted using three b = 0 images and 18 DWIs (a, iv), supervised learning denoised data using MU-Net with the ground-truth data as the training target (a, v), subset 1 of SDnDTI-denoised data (d, i) (i.e., the raw DWI (a, ii) denoised by SDnDTI), subset 2 of SDnDTI-denoised data (d, ii) (i.e., the synthesized DWI (a, iii) denoised by SDnDTI), the average of all three subsets of SDnDTI-denoised data (d, iii), BM4D-denoised data (d, iv), and AONLM-denoised data (d, v), along with a region of interest in the deep white matter with fine textures (yellow box in a, i) displayed in enlarged views (rows b, e) and residual images comparing to the ground-truth DWI (rows c, f). The mean absolute error (MAE), peak signal-to-noise ratio (PSNR) and the structural similarity index (SSIM) of different images comparing to the ground-truth DWI are used to quantify image similarity compared to the ground truth.

Figure 3 demonstrates that the denoised DWIs from SDnDTI show significantly improved SNR and quality (results for b = 0 image available in Supplementary Fig. S2). First, the DWI from subset 2 of SDnDTI input data (Fig. 3, a–c, iii) that was synthesized from the diffusion tensor generated using 6 DWIs only exhibited slightly higher noise level than that of the raw acquired DWI from subset 1 of SDnDTI input data (Fig. 3, a–c, ii), with comparable image similarity comparing to the ground-truth DWI (0.033 vs. 0.031 MAE, 27.10 dB vs 27.82 dB PSNR, and 0.875 vs 0.891 SSIM) due to the use of optimized diffusion-encoding directions, which did not make the subsequent denoising task more challenging. Consequently, the SDnDTI-denoised raw DWI (Fig. 3, d–f, i) and SDnDTI-denoised synthesized DWI (Fig. 3, d–f, ii) were indeed visually and quantitatively similar (0.017 vs. 0.017 MAE, 32.51 dB vs 32.32 dB, and 0.960 vs 0.961), as well as similar to the target DWI during the SDnDTI training that was synthesized from the diffusion tensor generated using all 18 DWI volumes (Fig. 3, a–c, iv, 0.018 MAE, 32.04 dB PSNR, 0.956 SSIM).

The final result of SDnDTI, i.e., the average of the three denoised DWIs from all three subsets, achieved further improved image quality (Fig. 3, d–f, iii, 0.012 MAE, 34.85 dB PSNR, 0.979 SSIM), which outperformed that from the target DWI during the SDnDTI training provided by the raw data (Fig. 3, a–c, iv), as well those from the raw DWI, the BM4D-denoised raw DWI (Fig. 3, d–f, iv, 0.022 MAE, 30.72 dB PSNR, 0.942 SSIM), and the AONLM-denoised raw DWI (Fig. 3, d–f, v, 0.023 MAE, 30.62 dB PSNR, 0.939 SSIM). The SDnDTI-denoised DWI also preserved more textural details around the internal capsule (Fig. 3, e, iii) comparing to the BM4D (Fig. 3, e, iv) and AONLM (Fig. 3, e, v) results, in which the stripe textures were blurred out. The resultant DWI from supervised denoising (Fig. 3, a–c, v) with the ground-truth DWI as the training target achieved the highest image similarity comparing to the ground-truth DWI (0.012 MAE, 36.43 dB PSNR, 0.982 SSIM), as expected, which slightly outperformed SDnDTI results. The residual maps between all denoised images and ground-truth images do not contain anatomical structure or biases reflecting the underlying anatomy (Fig. 3, rows c, f).

| | | a | b | c | d | e | f | g | h | i | j | k | l |
|---|---|---|---|---|---|---|---|---|---|---|---|---|---|
| | | Raw | Synth. (subset 1) | Synth. (subset 2) | Synth. (subset 3) | Synth. (18 DWIs) | Super-vised | SDnDTI (subset 1) | SDnDTI (subset 2) | SDnDTI (subset 3) | SDnDTI (average) | BM4D | AONLM |
| MAE | b = 0 | | 0.017 ± 0.00097 | 0.015 ± 0.00098 | 0.015 ± 0.00088 | 0.011 ± 0.00087 | 0.0099 ± 0.00073 | 0.014 ± 0.00091 | 0.013 ± 0.0087 | 0.013 ± 0.00078 | 0.01 ± 0.00082 | 0.014 ± 0.00063 | 0.013 ± 0.0007 |
| | DWI | 0.033 ± 0.0021 | 0.034 ± 0.0021 | 0.035 ± 0.0021 | 0.035 ± 0.0022 | 0.019 ± 0.0012 | 0.012 ± 0.00072 | 0.018 ± 0.0011 | 0.018 ± 0.00099 | 0.018 ± 0.0011 | 0.012 ± 0.00073 | 0.023 ± 0.0012 | 0.024 ± 0.0012 |
| PSNR (dB) | b = 0 | | 32.19 ± 0.62 | 32.94 ± 0.58 | 33.41 ± 0.6 | 35.31 ± 0.75 | 36.54 ± 0.76 | 33.18 ± 0.69 | 34.00 ± 0.6 | 34.35 ± 0.63 | 35.85 ± 0.77 | 33.67 ± 0.46 | 34.12 ± 0.55 |
| | DWI | 27.13 ± 0.57 | 26.89 ± 0.61 | 26.34 ± 0.61 | 26.45 ± 0.71 | 31.55 ± 0.8 | 35.83 ± 0.54 | 31.87 ± 0.86 | 31.68 ± 0.93 | 31.70 ± 1.09 | 34.50 ± 1.32 | 30.23 ± 0.47 | 30.16 ± 0.46 |
| SSIM | b = 0 | | 0.96 ± 0.0042 | 0.97 ± 0.004 | 0.97 ± 0.0039 | 0.99 ± 0.0023 | 0.99 ± 0.0019 | 0.98 ± 0.0029 | 0.98 ± 0.0025 | 0.98 ± 0.0024 | 0.99 ± 0.0018 | 0.98 ± 0.002 | 0.98 ± 0.0022 |
| | DWI | 0.88 ± 0.013 | 0.88 ± 0.013 | 0.86 ± 0.013 | 0.87 ± 0.013 | 0.95 ± 0.0054 | 0.98 ± 0.0023 | 0.96 ± 0.0047 | 0.96 ± 0.0047 | 0.96 ± 0.0047 | 0.98 ± 0.0018 | 0.94 ± 0.0056 | 0.94 ± 0.0058 |

**Table 1. Image similarity.** The group mean (± group standard deviation) across the 20 HCP subjects of the mean absolute error (MAE), peak signal-to-noise ratio (PSNR) and structural similarity index measure (SSIM) of b = 0 images and the diffusion-weighted image (DWI) shown in Figure 3 from the raw data (a), each subset of SDnDTI input data (b–d) (a single raw b = 0 image or DWIs synthesized from the diffusion tensor fitted using three b = 0 images and six DWIs), the averaged b = 0 image or DWIs synthesized from the diffusion tensor fitted using three b = 0 images and 18 DWIs (f), the raw data denoised by supervised learning with the ground-truth images as the training target (i.e., supervised-denoising) (f), each subset of SDnDTI-denoised data (g–i), the average of all three subsets of SDnDTI-denoised data (j), and the raw data denoised by BM4D (k) and AONLM (l) comparing to the ground-truth images generated from the tensor fitted using all 18 b = 0 and 90 DWIs.

The group mean (± the group standard deviation) across the 20 HCP subjects of the MAE, PSNR and SSIM between b = 0 images and the shown DWI in Figure 3 from different methods and ground-truth images were quantified in Table 1 (results for all DWIs available in Supplementary Tables 1–3, in which the shown DWI in Figure 3 is listed as "DWI 2"). The MAE of SDnDTI-denoised DWI was approximately one third of that of the raw DWI (0.012±0.00073 vs. 0.033±0.0021), two thirds of that of the target DWI during the SDnDTI training (0.019±0.0012), half of those of BM4D (0.023±0.0012) and AONLM results (0.024±0.0012), and equivalent to that of the supervised denoising results (0.012±0.00072). The group mean (± the group standard deviation) of the PSNR of SDnDTI-denoised DWI was approximately 7 dB higher than that of the raw DWI (34.50±1.32 dB vs. 27.13±0.57 dB), 3 dB higher than that of the target DWI during the SDnDTI training (31.55±0.92 dB), 4 dB higher than those of BM4D (30.23±0.47 dB) and AONLM results (30.16±0.46 dB), and 1.3 dB lower than that of the supervised denoising results (35.83±0.54 dB). The group mean (± the group standard deviation) of the SSIM of SDnDTI results was about 0.1 higher than that of the raw DWI (0.98±0.0024 vs. 0.88±0.013), 0.03 higher than that of the target

DWI during the SDnDTI training (0.95±0.0054), 0.04 higher than those of BM4D (0.94±0.0056) and AONLM results (0.94±0.0058), and equivalent to that of the supervised denoising results (0.98±0.0023).

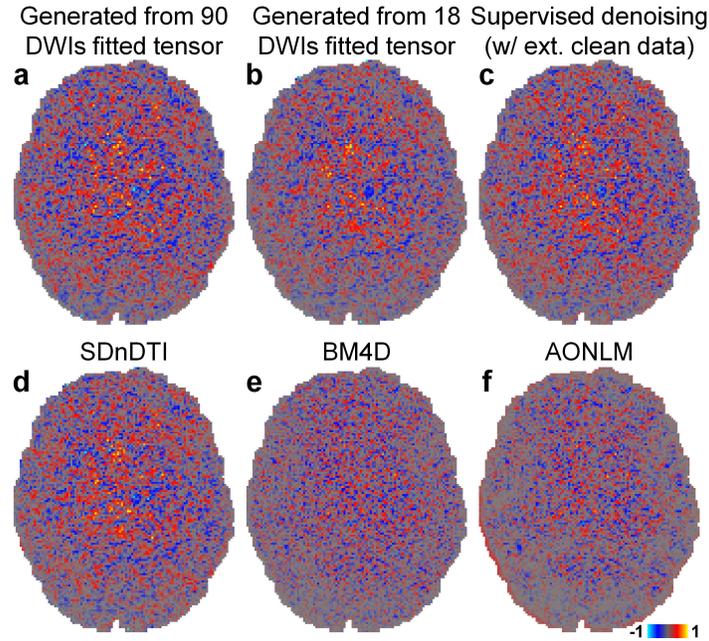

**Figure 4. Estimated noise.** Maps of the difference between the raw acquired diffusion-weighted image (DWI) sampled approximately along the superior-inferior direction (i.e., [-0.18, 0.26, -0.95]) shown in Figure 3 and the DWIs from different methods (i.e., the estimated noise) of a representative HCP subject.

Figure 4 demonstrates the capability of SDnDTI for estimating noise. The noise estimated by different methods (i.e., the residual maps between the acquired DWI and the denoised DWI) did not contain any noticeable anatomical structure or biases reflecting the underlying anatomy. The estimated noise maps of the synthesized DWI from the tensor fitted using 18 DWIs (Fig. 4b), the supervised denoising results (Fig. 4c) and the SDnDTI results (Fig. 4d) were visually more similar to the noise map from the ground-truth DWI (Fig. 4a) than those from the BM4D (Fig. 4e) and AONLM (Fig. 4f) results.

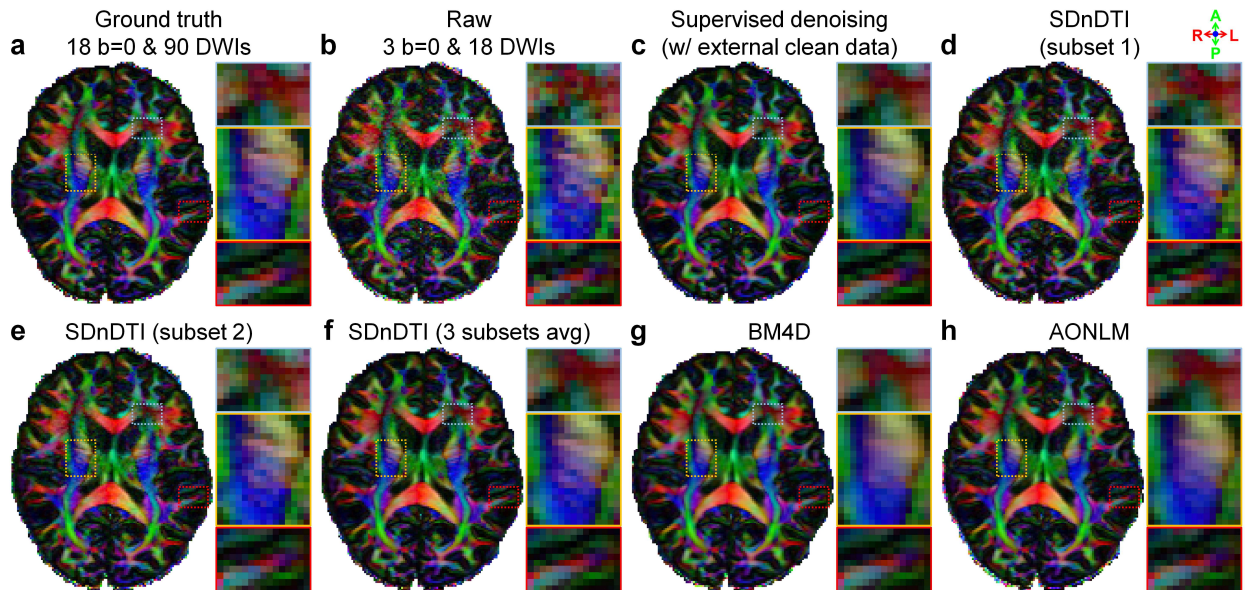

**Figure 5. Structure mapping.** Fractional anisotropy (FA) maps color encoded by the primary eigenvector (red: left–right; green: anterior–posterior; blue: superior–inferior) derived from the diffusion tensors fitted using all 18 b = 0 and 90 diffusion-weighted images (DWIs) (ground truth, a), raw data consisting of three b = 0 and 18 DWIs (b), the raw data denoised by supervised learning with the ground-truth images as the training target (i.e., supervised denoising) (c), the subset 1 of SDnDTI-denoised data (d), the subset 2 of SDnDTI-denoised data (e), the average of all three subsets of SDnDTI-denoised data (f), and the raw data denoised by BM4D (g) and AONLM (h) from a representative HCP subject. Three regions of interest in the deep white matter (yellow boxes) and sub-cortical white matter surrounded by gray matter (red boxes) or with intersecting fiber tracts (blue boxes) are displayed in enlarged views.

Figure 5 shows the ability of SDnDTI to recover detailed anatomical information from the noisy inputs as mapped in primary eigenvector V1-encoded FA maps. The FA maps from a single subset of SDnDTI-denoised data (Fig. 5d, e) were substantially less noisy compared to the map derived from the raw data (Fig. 5b). The FA map from the average of three subsets of SDnDTI-denoised data (Fig. 5f) further improved upon the maps from each single subset data (Fig. 5d, e) and was visually similar to the map from the supervised denoising (Fig. 5c). The SDnDTI map (Fig. 5f) was slightly blurred compared to the ground-truth map (Fig. 5a), but sharper than the map derived from BM4D (Fig. 5g) and AONLM (Fig. 5h), which was clearly depicted in the gray matter bridges that span the internal capsule with characteristic stripes (Fig. 5f–h, yellow boxes). These textures were buried in noise in the map derived from raw data (Fig. 5b, yellow box). Because of the blurring, the FA of the cortical gray matter in the BM4D- and AONLM-denoised results also significantly reduced (the green contour surrounding the gyrus in Fig. 5g, h, red boxes), which was preserved more in the SDnDTI-denoised results (Fig. 5f, red box).

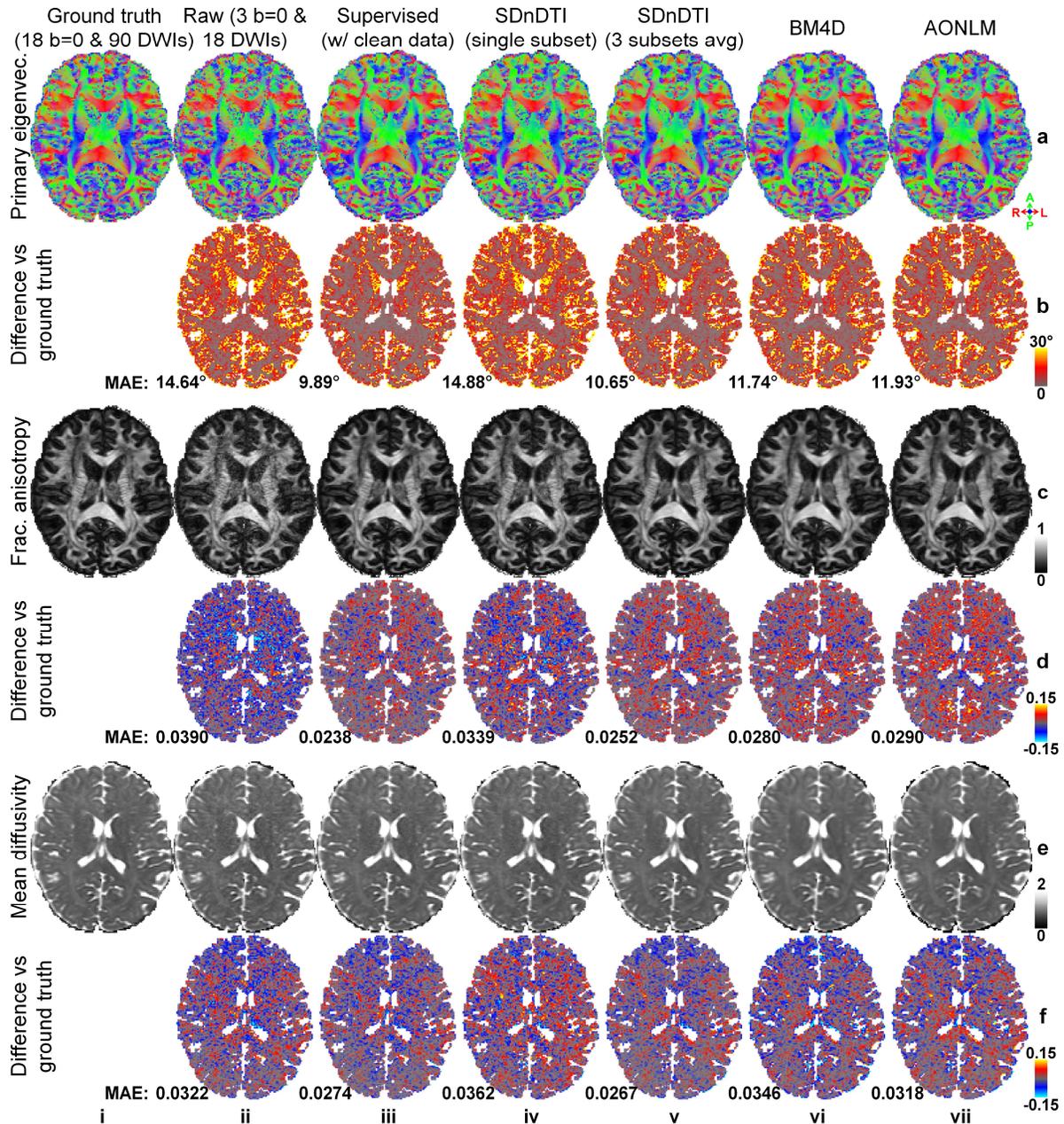

**Figure 6. DTI metrics.** Maps of color-encoded primary eigenvector (red: left–right; green: anterior–posterior; blue: superior–inferior) (row a), fractional anisotropy (row c) and mean diffusivity (row e) derived from the diffusion tensors fitted using all 18 b = 0 and 90 diffusion-weighted images (DWIs) (ground truth, column i), raw data consisting of three b = 0 and 18 DWIs (column ii), the raw data denoised by supervised learning with the ground-truth images as the training target (i.e., supervised denoising) (column iii), the subset 2 of SDnDTI-denoised data (column iv), the average of all three subsets of SDnDTI-denoised data (column v), and the raw data denoised by BM4D (column vi) and AONLM (column vi), and their residual maps (rows b, d, f) compared to the ground-truth maps from a representative HCP subject. The mean absolute difference (MAD) of each map compared to the ground truth within the brain (excluding the cerebrospinal fluid) is displayed at the bottom of the residual map. The unit of the diffusivity is μm²/ms.

| | HCP Data | Primary eigenvector (°) | Fractional anisotropy | Mean diffusivity (μm²/ms) | Axial diffusivity (μm²/ms) | Radial diffusivity (μm²/ms) |
|---|---|---|---|---|---|---|
| a | Raw | 15.48 ± 1.05 | 0.042 ± 0.0031 | 0.033 ± 0.0028 | 0.057 ± 0.0041 | 0.037 ± 0.003 |
| b | Raw (subset 1) | 24.90 ± 1.25 | 0.086 ± 0.0069 | 0.043 ± 0.0030 | 0.10 ± 0.0071 | 0.056 ± 0.0044 |
| c | Raw (subset 2) | 24.92 ± 1.32 | 0.086 ± 0.0067 | 0.043 ± 0.0032 | 0.10 ± 0.0075 | 0.056 ± 0.0043 |
| d | Raw (subset 3) | 24.86 ± 1.34 | 0.085 ± 0.0069 | 0.043 ± 0.0033 | 0.10 ± 0.0079 | 0.056 ± 0.0043 |
| e | Supervised | 10.52 ± 0.75 | 0.025 ± 0.0014 | 0.028 ± 0.0023 | 0.040 ± 0.0025 | 0.030 ± 0.0024 |
| f | SDnDTI (subset 1) | 15.73 ± 1.05 | 0.036 ± 0.0021 | 0.039 ± 0.0029 | 0.058 ± 0.0034 | 0.042 ± 0.0030 |
| g | SDnDTI (subset 2) | 15.99 ± 1.11 | 0.036 ± 0.0021 | 0.037 ± 0.0027 | 0.057 ± 0.0034 | 0.041 ± 0.0027 |
| h | SDnDTI (subset 3) | 15.80 ± 1.09 | 0.035 ± 0.0019 | 0.037 ± 0.0027 | 0.056 ± 0.0033 | 0.040 ± 0.0027 |
| i | SDnDTI (average) | 11.20 ± 0.81 | 0.026 ± 0.0015 | 0.028 ± 0.0025 | 0.041 ± 0.0027 | 0.030 ± 0.0025 |
| j | BM4D | 12.46 ± 0.89 | 0.029 ± 0.0016 | 0.036 ± 0.0027 | 0.051 ± 0.0031 | 0.037 ± 0.0026 |
| k | AONLM | 12.64 ± 0.87 | 0.030 ± 0.0016 | 0.033 ± 0.0027 | 0.050 ± 0.0032 | 0.035 ± 0.0026 |

**Table 2. Errors of DTI metrics.** The group mean (± group standard deviation) across the 20 HCP subjects of the mean absolute difference (MAD) of different DTI metrics derived from the raw data consisting of three b = 0 and 18 diffusion-weighted images (DWIs) (a), each subset of SDnDTI input data (b–d), the raw data denoised by supervised learning with the ground-truth images as the training target (i.e., supervised-denoising) (e), each subset of SDnDTI-denoised data (f–h), the average of all three subsets of SDnDTI-denoised data (i), and the raw data denoised by BM4D (j) and AONLM (k) comparing to the ground-truth DTI metrics derived from 18 b = 0 and 90 DWIs.

The difference of five common DTI metrics, including V1, FA, MD, AD, RD between the results derived from different methods and ground-truth data is displayed for a representative subject (Fig. 6, Supplementary Fig. S3) and quantified for 20 HCP subjects (Table 2). The group means of the MAD derived from each subset of SDnDTI-denoised data (Fig. 6, iv, Table 2, f–h) remarkably improved upon the SDnDTI inputs (Table 2, b–d), which was similar to those from the raw data (Fig. 6, ii, Table 2, a), as expected. Because of the averaging, the group mean of the MAD derived from the final results of SDnDTI (Fig. 6, v, Table 2, i) substantially outperformed those from the raw data (Fig. 6, ii, Table 2, a), as well as were superior to those from BM4D-denoised (Fig. 6, vi, Table 2, j) and AONLM-denoised (Fig. 6, vii, Table 2, k) raw data. The group means of the MAD of BM4D- and AONLM-denoised raw data were in general very similar. The supervised-denoising (Fig. 6, iii, Table 2, e) achieved the lowest MAD for all DTI metrics, which were marginally lower than those from SDnDTI results (i.e., 0.025±0.0014 vs. 0.026±0.0015 for FA, 0.028±0.0023 μm²/ms vs. 0.028±0.0023 μm²/ms for MD, 0.040±0.0025 μm²/ms vs. 0.041±0.0027 μm²/ms for AD, and 0.030±0.0024 μm²/ms vs. 0.030±0.0025 μm²/ms for RD), except for the MAD for primary eigenvector (i.e., 10.52°±0.75° vs. 11.20°±0.81°).

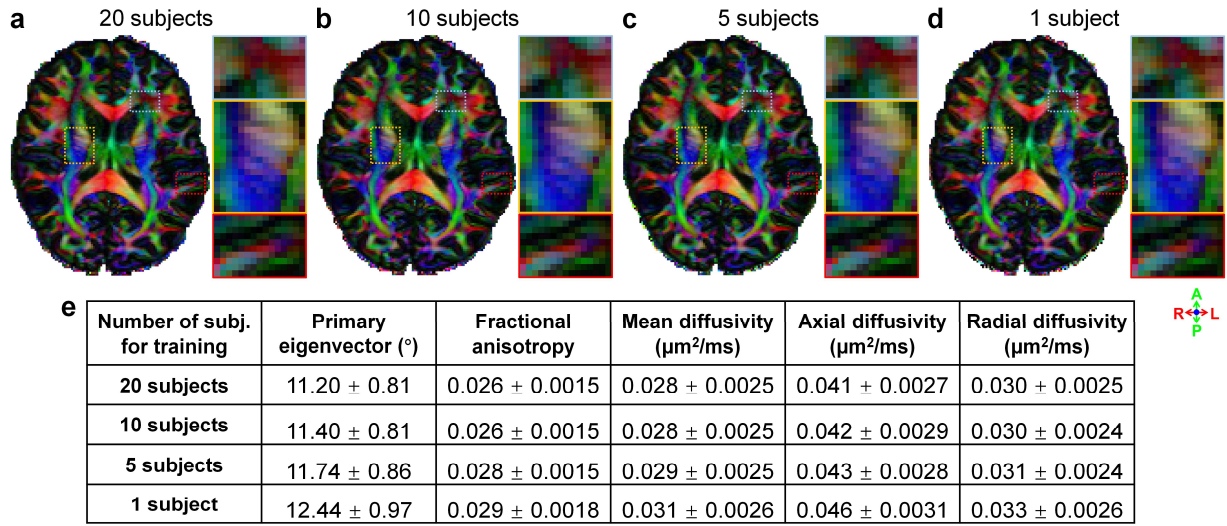

**Figure 7. Effects of training data.** Fractional anisotropy maps color encoded by the primary eigenvector (red: left–right; green: anterior–posterior; blue: superior–inferior) derived from the raw data consisting 3 b = 0 and 18 DWIs denoised by SDnDTI trained using data of 20 subjects (a), 10 subjects (b), 5 subjects (c) and 1 subject (d) from the HCP. Three regions of interest in the deep white matter (yellow boxes) and sub-cortical white matter surrounded by gray matter (red boxes) or with intersecting fiber tracts (blue boxes) are displayed in enlarged views. The table lists the group mean (± group standard deviation) across the 20 HCP subjects of the mean absolute difference (MAD) of different DTI metrics comparing to the ground-truth DTI metrics derived from 18 b = 0 and 90 DWIs.

| Number of subj. for training | Primary eigenvector (°) | Fractional anisotropy | Mean diffusivity ($\mu m^2$/ms) | Axial diffusivity ($\mu m^2$/ms) | Radial diffusivity ($\mu m^2$/ms) |
|---|---|---|---|---|---|
| 20 subjects | 11.20 + 0.81 | 0.026 + 0.0015 | 0.028 ± 0.0025 | 0.041 + 0.0027 | 0.030 + 0.0025 |
| 10 subjects | 11.40 ± 0.81 | 0.026 ± 0.0015 | 0.028 ± 0.0025 | 0.042 ± 0.0029 | 0.030 ± 0.0024 |
| 5 subjects | 11.74 ± 0.86 | 0.028 ± 0.0015 | 0.029 ± 0.0025 | 0.043 ± 0.0028 | 0.031 ± 0.0024 |
| 1 subject | 12.44 ± 0.97 | 0.029 ± 0.0018 | 0.031 ± 0.0026 | 0.046 ± 0.0031 | 0.033 ± 0.0026 |

The denoising performance of SDnDTI using different number of training subjects is depicted in Figure 7. The primary eigenvector V1-encoded FA maps were visually very similar without noticeable difference. Even when the MU-Net of SDnDTI was trained on the data of each single subject, SDnDTI could still preserve image sharpness, such as the strip texture spanning the internal capsule (Fig. 7d), superior to BM4D and AONLM (Fig. 5g, h). Quantitatively, the group means of the MAD in DTI metrics were the lowest if the data of the 20 HCP subjects were jointly denoised, in which case there were plenty of training data for optimizing the MU-Net of SDnDTI. When the number of training subjects reduced from 20 to 10, the group means of the MAD in DTI metrics only marginally increased, presumably because the data from 10 subjects were still sufficient to train the MU-Net of SDnDTI. The denoising performance decreased more rapidly from 10 to 5 subjects, and 5 to 1 subject, especially for the primary eigenvector. Even when the MU-Net of SDnDTI was trained on the data of each single subject for denoising, the group means of the MAD in DTI metrics of SDnDTI were still slightly lower than those from BM4D and AONLM (e.g., 12.44°±0.97° vs. 12.46°±0.89° and 12.64°±0.87° for primary eigenvector, 0.031±0.0026 $\mu m^2$/ms vs. 0.036±0.0027 $\mu m^2$/ms and 0.033±0.0027 $\mu m^2$/ms for MD).

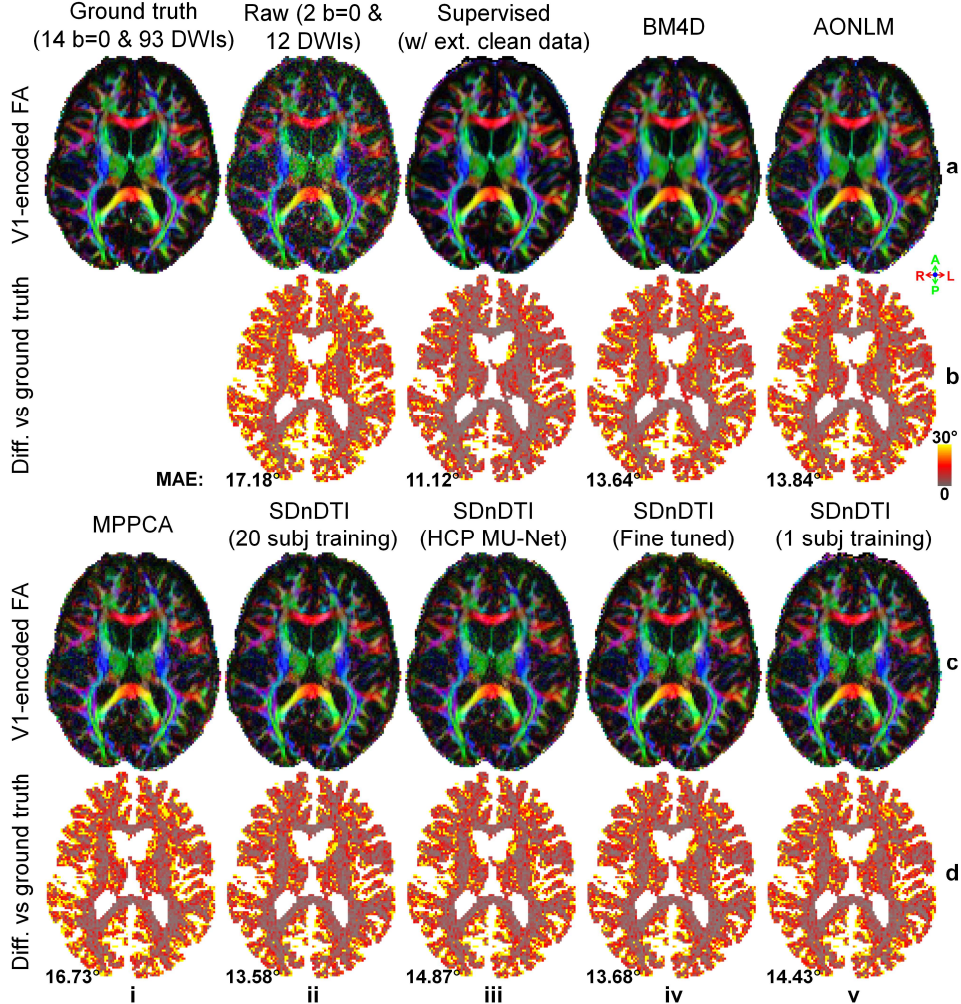

**Figure 8. Primary eigenvector.** Maps of color-encoded primary eigenvector (red: left–right; green: anterior–posterior; blue: superior–inferior) modulated by the fractional anisotropy (row a, c) derived from the diffusion tensors fitted using all 14 b = 0 and 93 diffusion-weighted images (DWIs) (ground truth, a, i), raw data consisting of two b = 0 and 12 DWIs (a, ii), the raw data denoised by supervised learning with the ground-truth DWIs as the training target (i.e., supervised denoising) (a, iii), BM4D (a, iv), AONLM (a, v) and MPPCA (c, i), and SDnDTI (c, ii–v), and their residual maps (rows b, d) compared to the ground-truth map from a representative HCP-A subject. SDnDTI results were generated by an MU-Net trained on the data from 20 HCP-A subjects (c, ii), an MU-Net trained on the data from 20 HCP subjects (c, iii), an MU-Net with parameters from the MU-Net trained on the data from 20 HCP subjects as initialization and further fine-tuned using the data of each HCP-A subject (c, iv), and an MU-Net trained on the data from the data of each HCP-A subject (c, v). The mean absolute difference (MAD) of each map compared to the ground truth within the brain (excluding the cerebrospinal fluid) is displayed at the bottom of the residual map.

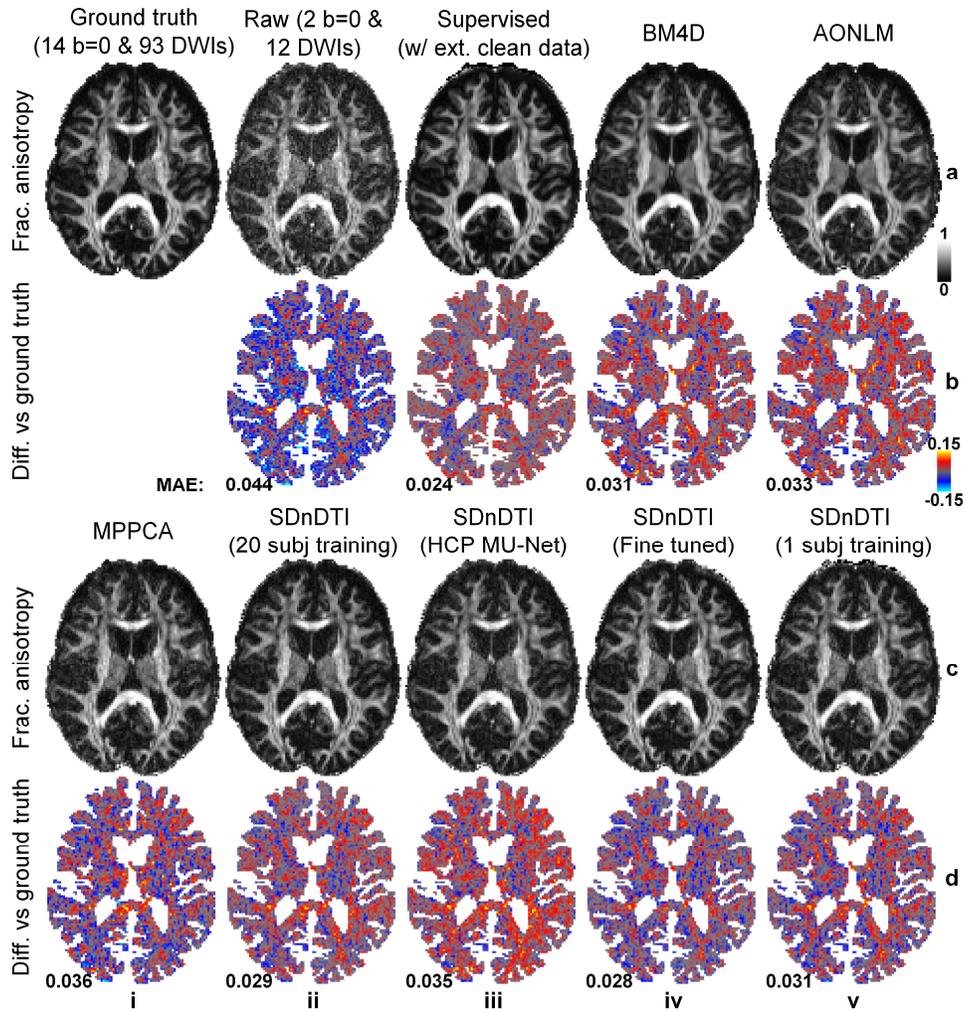

**Figure 9. Fractional anisotropy.** Maps of fractional anisotropy (row a, c) derived from the diffusion tensors fitted using all 14 b = 0 and 93 diffusion-weighted images (DWIs) (ground truth, a, i), raw data consisting of two b = 0 and 12 DWIs (a, ii), the raw data denoised by supervised learning with the ground-truth DWIs as the training target (i.e., supervised denoising) (a, iii), BM4D (a, iv), AONLM (a, v) and MPPCA (c, i), and SDnDTI (c, ii–v), and their residual maps (rows b, d) compared to the ground-truth map from a representative HCP-A subject. SDnDTI results were generated by an MU-Net trained on the data from 20 HCP-A subjects (c, ii), an MU-Net trained on the data from 20 HCP subjects (c, iii), an MU-Net with parameters from the MU-Net trained on the data from 20 HCP subjects as initialization and further fine-tuned using the data of each HCP-A subject (c, iv), and an MU-Net trained on the data from the data of each HCP-A subject (c, v). The mean absolute difference (MAD) of each map compared to the ground truth within the brain (excluding the cerebrospinal fluid) is displayed at the bottom of the residual map.

| | HCP-A data | Primary eigenvector (°) | Fractional anisotropy | Mean diffusivity ($\mu m^2/ms$) | Axial diffusivity ($\mu m^2/ms$) | Radial diffusivity ($\mu m^2/ms$) |
|---|---|---|---|---|---|---|
| a | Raw | 16.17 ± 0.72 | 0.043 ± 0.0029 | 0.025 ± 0.0023 | 0.054 ± 0.0035 | 0.030 ± 0.0024 |
| b | Supervised | 10.14 ± 0.56 | 0.023 ± 0.00092 | 0.020 ± 0.0023 | 0.032 ± 0.0021 | 0.023 ± 0.0023 |
| c | BM4D | 12.78 ± 0.58 | 0.031 ± 0.0014 | 0.032 ± 0.0040 | 0.049 ± 0.0040 | 0.034 ± 0.0034 |
| d | AONLM | 13.04 ± 0.61 | 0.033 ± 0.0016 | 0.027 ± 0.0027 | 0.046 ± 0.0028 | 0.030 ± 0.0023 |
| e | MPPCA | 15.71 ± 0.69 | 0.036 ± 0.0019 | 0.025 ± 0.0024 | 0.047 ± 0.0028 | 0.028 ± 0.0024 |
| f | SDnDTI (20 subj) | 12.63 ± 0.61 | 0.028 ± 0.0014 | 0.022 ± 0.0022 | 0.039 ± 0.0021 | 0.024 ± 0.0022 |
| g | SDnDTI (HCP MU-Net) | 14.06 ± 0.58 | 0.036 ± 0.0014 | 0.023 ± 0.0023 | 0.045 ± 0.0019 | 0.027 ± 0.0022 |
| h | SDnDTI (fine tuned) | 12.66 ± 0.70 | 0.028 ± 0.0016 | 0.022 ± 0.0023 | 0.039 ± 0.0025 | 0.025 ± 0.0023 |
| i | SDnDTI (1 subj) | 13.65 ± 0.65 | 0.031 ± 0.0015 | 0.024 ± 0.0024 | 0.043 ± 0.0025 | 0.027 ± 0.0023 |

**Table 3. Errors of DTI metrics.** The group mean (± group standard deviation) across the 20 HCP-A subjects of the mean absolute difference (MAD) of different DTI metrics derived from the raw data consisting of two b = 0 and 12 diffusion-weighted images (DWIs) (a), the raw data denoised by supervised learning with the ground-truth DWIs as the training target (i.e., supervised denoising) (b), BM4D (c), AONLM (d), MPPCA (e), and SDnDTI (f–i) comparing to the ground-truth DTI metrics derived from 14 b = 0 and 93 DWIs. SDnDTI results were generated by an MU-Net trained on the data from 20 HCP-A subjects (f), an MU-Net trained on the data from 20 HCP subjects (g), an MU-Net with parameters from the MU-Net trained on the data from 20 HCP subjects as initialization and further fine-tuned using the data of each HCP-A subject (h), and an MU-Net trained on the data from the data of each HCP-A subject (i).

The difference of five common DTI metrics, including V1, FA, MD, AD, RD between the results derived from different approaches and ground-truth data is displayed for a representative HCP-A subject (Fig. 8 for V1, Fig. 9 for FA, Supplementary Figs. S4–S6 for MD, AD and RD) and quantified for the 20 HCP-A subjects (Table 3). SDnDTI substantially improved the SNR of the V1 and FA maps from the raw data (Figs. 8, 9, a, ii, vs. c, ii). The FA maps from supervised denoising (Fig. 9, a, iii), SDnDTI (Fig. 9, c, ii) and MPPCA (Fig. 9, c, i) appear sharper than those from BM4D (Fig. 9, a, iv) and AONLM (Fig. 9, a, v) and are visually more similar to the ground-truth map (Fig. 9, a, i).

Quantitatively, the group means of the MAD from SDnDTI (Table 3f) were substantially lower than those from the raw data (Table 3, a), as well as were superior to those from BM4D (Table 3, c), AONLM (Table 3, d) and MPPCA (Table 3, e), especially for MD, AD and RD for BM4D and AONLM and primary eigenvector, FA, AD and RD for MPPCA. As expected, the group means of the MAD of supervised denoising were the lowest (Table 3, a), which were slightly lower than those from SDnDTI for scalar metrics FA, MD, AD and RD while had greater advantage comparing to those from SDnDTI for the primary eigenvector. The MU-Net of the SDnDTI generalized to different data reasonably well. When applying the MU-Net of SDnDTI trained on the data from HCP subjects directly to the data from HCP-A subjects acquired with very different hardware systems and protocols, the resultant maps were cleaner than those from the raw data (Figs. 8, 9, a, ii vs. c, iii), with the group means of the MAD (Table 3, g) lower than those from the raw data (Table 3, a) and MPPCA (Table 3, e), but higher than those from BM4D (Table 3, c), AONLM (Table 3, d) and SDnDTI trained on data from the 20 HCP-A subjects (Table 3, f). If the HCP MU-Net was further fine-tuned using the data of each single HCP-A subject, the denoising performance of the fine-tuned MU-Net was equivalent to the one optimized using much more training data from 20 HCP-A subjects (Figs. 8, 9, c, ii, vs. c, iv, Table 3, f vs. h). In contrast, the denoising performance of the MU-Nets trained from random initialization on each single HCP-A subject was slightly inferior to that of the fine-tuned ones (Table 3, h vs. i), which still outperformed the raw data (Table 3, a) and MPPCA (Table 3, e).

**Discussion**

In this study, we have developed a self-supervised deep learning approach called SDnDTI for denoising DTI data that does not require external high-SNR data for training. SDnDTI relies on the "first denoising then averaging" mechanism by first denoising each single image volume of the multi-directional DTI data with the averaged image volume as the target using CNNs and then averaging multiple denoised results. The performance of SDnDTI is systematically evaluated in terms of the quality of output images and DTI metrics, as well as compared to supervised learning based denoising and conventional state-of-the-art denoising algorithms BM4D, AONLM and MPPCA on two different datasets provided by HCP and HCP-A. SDnDTI-denoised images preserve textural details and are sharper than those from BM4D and AONLM as well as similar to the ground truth with low MAEs of ~0.01 for b = 0 images and ~0.012 for DWIs, high PSNR of ~36 dB for b = 0 images and ~35 dB for DWIs, and high SSIM of ~0.99 for b = 0 images and ~0.98 for DWIs. SDnDTI derived images and DTI metrics are comparable to those from supervised denoising, substantially outperform those from the raw data, and are superior to those from BM4D, AONLM and MPPCA. SDnDTI is capable to generalize to different datasets and benefits from further fine tuning and more training data when the data of numerous subjects are jointly denoised.

The concept of "first denoising then averaging" rather than "directly averaging" could be applied to any applications in which multiple measurements, such as many repetitions of $T_1$-weighted data at high isotropic spatial resolution[77,109] and numerous DWI volumes sampled along many directions in DTI, are acquired. The efficacy of this concept relies on the superior performance of deep learning-based denoising using CNNs, which are capable to map the noisier image data to the cleaner image data without compromising image quality (Figure 3, Supplementary Figure S2, Table 1, Supplementary Table 1–3). The superior performance is due to the use of residual learning and deep 3D CNNs. On the one hand, residual learning (i.e., learning the residuals between the input noisier image data and the target high-SNR image data) not only boosts the CNN performance[59,89,110,111] since the CNN only needs to synthesize the high-frequency information, but also preserves image sharpness and textual details. On the other hand, deep 3D CNNs (i.e., 10-layer 3D MU-Net in SDnDTI) can fully exploit the redundant information contained in the data. DTI is the imaging modality that benefits the most from the "first denoising then averaging" concept since numerous DWI volumes are acquired by nature. CNNs can map a single DWI volume to the cleaner target generated using much more data (e.g., 18 or 12 DWI volumes for HCP and HCP-A data in our study). If only two or three repetitions of image data (e.g., $T_1$-weighted or $T_2$-weighted) are acquired, the denoising effect is not as strong as for the multi-directional diffusion data.

Implementing the "first denoising then averaging" concept for denoising multi-directional DWI volumes leverages domain knowledge of diffusion MRI physics. The challenge lies in the fact that DWI volumes in DTI are sampled along uniformly distributed directions thus exhibiting different image contrast, while this concept requires several repetitions of DWI volumes with identical image contrast but different noise observations. SDnDTI addresses this challenge by transforming DWIs sampled along one set of directions to the other set through the diffusion tensor model (Equations 1, 2), i.e., first fitting a tensor model using DWIs along a set of directions and then synthesizing DWI volumes along another set of directions. In this way, the DWI volumes from all subsets can be transformed to the same diffusion-encoding directions, and the target DWI volumes with higher SNR can be also synthesized using the tensor fitted using all acquired data along these directions. To avoid the loss of angular sampling coverage, SDnDTI synthesizes DWIs along all acquired directions rather than fewer directions as in each subset of data.

The diffusion-encoding directions in each subset of DTI data must be carefully selected such that the noise is not amplified during the image transformation process. Otherwise, the subsequent denoising task becomes more difficult and the denoising results cannot match the target DWIs even using a CNN. SDnDTI adopts the uniform encoding directions from the DSM scheme that minimize the condition number of the diffusion tensor transformation matrix to 1.3228, which successfully suppress noise amplification during the transformation. The transformed images are only slightly noisier than the raw acquired images (Fig. 3, a–c, ii, Table 2, a–d). Since the HCP and HCP-A data are pre-required and thus the directions used in this study are only approximately as designed in the DSM scheme (i.e., the condition number of the diffusion tensor transformation matrix is ~1.6). We expect SDnDTI performance to be increased if the actual DSM directions could be used for the data acquisition. Moreover, SDnDTI uses 6 DWI volumes for each subset because the HCP and HCP-A date are not very noisy. For extremely noisy DTI data such as those from sub-millimeter isotropic resolution, data along much more directions are often acquired (e.g., two repetitions of 256 directions at 1-mm isotropic spatial resolution[17]) and more DWI volumes (e.g., 20, 30 DSM directions) should be assigned for each subset for robust tensor fitting and image transformation.

It is beneficial to jointly denoise the DTI data of all subjects in a study using SDnDTI because the CNN performance improves as the amount of training data increases (Figure 7). It is not trivial to theoretically determine the required number of subjects for training, which depends on the number of parameters of the CNN (~12 million for the adopted 3D MU-Net) and the information contained in the image data of each subject (influenced by factors such as the brain size, image resolution and the number of DWI volumes in a DTI dataset). Empirically, the performance of SDnDTI trained and validated using data from 10 subjects from HCP is almost identical to that of using data from 20 subjects while decreases if using data from 5 subjects (Figure 7), suggesting that data from at least 10 subjects are required to optimize the MU-Net of SDnDTI given the HCP imaging parameters (e.g., 1.25 mm isotropic resolution) and DTI protocol (e.g., 3 b = 0 image volumes and 18 DWI volumes). On the other hand, a CNN with less parameters (e.g., shallower or with less kernels at each layer) can be adopted, which might be well trained using limited data and be more effective, but the trade-off between the CNN parameter number and the subject number needs to be determined empirically.

A preferable approach to account for the limited training data is to fine-tune parameters of the SDnDTI CNN pre-trained using big data provided by large-scale neuroimaging studies such as HCP. Our experiments show that SDnDTI results from the MU-Nets trained and validated using data from each single HCP-A subject cannot compare to those from BM4D and AONLM, even though they outperform those from the raw data, MPPCA and the MU-Nets trained and validated on the HCP data (Table 3). This is presumably due to insufficient training data, since SDnDTI results from the MU-Net trained and validated using data from 20 HCP-A subjects indeed improve and outperform those from BM4D and AONLM (Table 3). However, further adapting parameters of the MU-Net trained and validated on the HCP data using the data from each single HCP-A subject substantially increase the quality of denoised results (Table 3), which outperforms that from all tested conventional denoising algorithms and is essentially identical to that from the MU-Net trained and validated using 20 HCP-A subjects. In addition to reducing the requirement for training data, fine-tuning also helps accelerate training convergence and reduce training time. Fine-tuning the HCP MU-Net using data from each single HCP-A subject only requires 8 to 34 minutes training and validation time, while training and validating the MU-Net with randomly initialized parameters on data from 20 HCP-A subjects take ~20 hours, even though they output identical results. We will make our codes for SDnDTI publicly available (https://github.com/qiyuantian/SDnDTI) which can be used to pre-train the CNN of SDnDTI using HCP data for fine-tuning, a recommended approach in practice, or train the CNN for supervised learning using HCP data for fine-tuning if additional high-quality data is available from a few subjects in some applications, after all the supervised denoising achieves the highest performance (Figs. 5, 6, 8, 9, Tables 2, 3).

**Summary**

This study presents a data-driven self-supervised deep learning-based denoising method entitled SDnDTI for DTI that does not require additional high-SNR data as the target for training. SDnDTI works by first denoising each single image volume of the multi-directional DTI data with the averaged image volume as the target using CNNs and then averaging multiple denoised results for recovering even higher SNR, a concept known as "first denoising then averaging". SDnDTI-denoised DWIs preserve image sharpness and textural details and are similar to the ground-truth DWIs with low MAEs of ~0.012, high PSNR of ~35 dB, and high SSIM of ~0.98. SDnDTI-denoised images and derived DTI metrics are comparable to results from supervised learning-based denoising that use ground-truth images as the training target, and are superior to results from the raw data, and BM4D-, AONLM- and MPPCA-denoised data. SDnDTI generalizes well to different datasets and fine-tuning parameters of the pre-trained CNN of SDnDTI further improves denoising performance as well as shortens training time. By excluding the need for external high-SNR data and the generalization of CNNs, SDnDTI increases the feasibility of deep learning and CNN-based denoising methods in a wider range of clinical and neuroscientific studies that benefit from faster DTI acquisition and improved DTI data quality.

**Acknowledgments**

This work was supported by the NIH Grants P41-EB015896, P41-EB030006, U01-EB026996, U01-EB025162, U01-AG052564, K23-NS096056, R01-EB017337, R01-MH111419, R01-EB028797, R03-EB031175, and an MGH Claflin Distinguished Scholar Award. The diffusion and anatomical data of 20 young healthy adults were provided by the Human Connectome Project, WU-Minn-Ox Consortium (Principal Investigators: David Van Essen and Kamil Ugurbil; U54-MH091657) funded by the 16 NIH Institutes and Centers that support the NIH Blueprint for Neuroscience Research; and by the McDonnell Center for Systems Neuroscience at Washington University. B.B. has provided consulting services to Subtle Medical.

**Data availability**

The diffusion and $T_1$-weighted MRI data of 20 subjects from the Human Connectome Project WU-Minn-Ox Consortium and are publicly available (https://www.humanconnectome.org). The diffusion and $T_1$-weighted MRI data from the Lifespan Human Connectome Project in Aging are publicly available (https://www.humanconnectome.org/study/hcp-lifespan-aging)

**Code availability**

The source codes of BM4D implemented using MATLAB are publicly available (https://www.cs.tut.fi/~foi/GCF-BM3D). The MATLAB-based software of AONLM is publicly available (https://sites.google.com/site/pierrickcoupe/softwares/denoising-for-medical-imaging/mri-denoising/mri-denoising-software). The source codes of MPPCA implemented using MATLAB are publicly available (https://github.com/NYU-DiffusionMRI/mppca_denoise). The source codes of SDnDTI implemented using MATLAB and Keras application programming interface will be made publicly available (https://github.com/qiyuantian/SDnDTI).

**Supplementary Information**

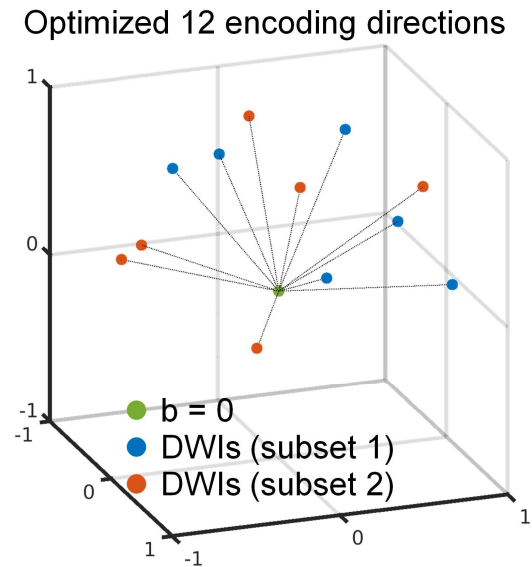

**Supplementary Figure S1. Optimized diffusion-encoding directions.** The 12 diffusion-encoding directions for the HCP-A data are optimized such that they can be divided into two subsets of 6 directions which minimize the condition number of the diffusion tensor transformation matrix, while are also uniformly distributed on a sphere.

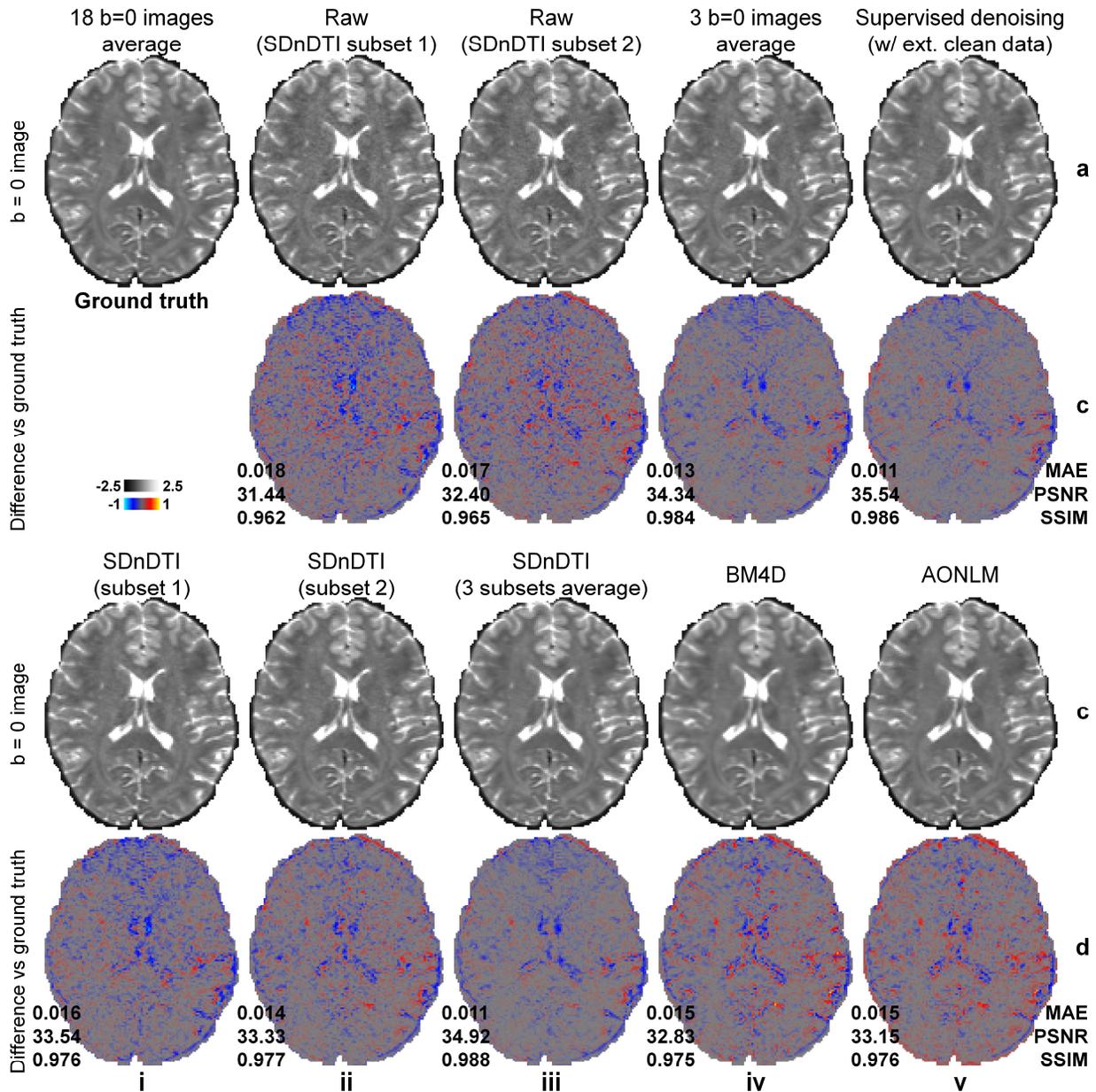

**Supplementary Figure S2. Denoised b = 0 images.** The average of all 18 b = 0 images (ground-truth b = 0 image) (a, i), a single raw b = 0 image from subset 1 of SDnDTI inputs (a, ii), another single raw b = 0 image from subset 2 of SDnDTI inputs (a, iii), the average of 3 b = 0 images (a, iv), the average of all three denoised b = 0 images using MU-Net with the ground-truth b = 0 image as the training target (a, v), the single b = 0 image (a, ii) denoised by SDnDTI (c, i), another single b = 0 image (a, iii) denoised by SDnDTI (c, ii), the average of all three SDnDTI-denoised b = 0 images (c, iii), BM4D-denoised b = 0 image (c, iv), and AONLM-denoised b = 0 image (c, v), along with residual images comparing to the ground-truth b = 0 image (rows b, d) from a representative HCP subject. The mean absolute error (MAE), peak signal-to-noise ratio (PSNR) and the structural similarity index (SSIM) of different images comparing to the ground-truth b = 0 image are used to quantify image similarity compared to the ground truth.

| Mean absolute error | b = 0 | DWI 1 | DWI 2 | DWI 3 | DWI 4 | DWI 5 | DWI 6 | DWI 7 | DWI 8 | DWI 9 | DWI 10 | DWI 11 | DWI 12 | DWI 13 | DWI 14 | DWI 15 | DWI 16 | DWI 17 | DWI 18 |
|---|---|---|---|---|---|---|---|---|---|---|---|---|---|---|---|---|---|---|---|
| Raw | 0.015 ± 0.00088 | 0.033 ± 0.0022 | 0.033 ± 0.0021 | 0.033 ± 0.0022 | 0.033 ± 0.0022 | 0.033 ± 0.0021 | 0.033 ± 0.0022 | 0.035 ± 0.0023 | 0.034 ± 0.0021 | 0.033 ± 0.0021 | 0.033 ± 0.0022 | 0.034 ± 0.0023 | 0.033 ± 0.0021 | 0.035 ± 0.0024 | 0.034 ± 0.0023 | 0.033 ± 0.0021 | 0.034 ± 0.0022 | 0.033 ± 0.0022 | 0.034 ± 0.0021 |
| Gen. from 6 DWIs fit tensor (subset 1) | 0.017 ± 0.00067 | 0.033 ± 0.0022 | 0.034 ± 0.0021 | 0.033 ± 0.0022 | 0.034 ± 0.0022 | 0.033 ± 0.0021 | 0.033 ± 0.0022 | 0.033 ± 0.0022 | 0.036 ± 0.0025 | 0.039 ± 0.0025 | 0.033 ± 0.0022 | 0.034 ± 0.0023 | 0.035 ± 0.0023 | 0.032 ± 0.0021 | 0.031 ± 0.002 | 0.033 ± 0.0021 | 0.041 ± 0.0029 | 0.033 ± 0.0022 | 0.039 ± 0.0027 |
| Gen. from 6 DWIs fit tensor (subset 2) | 0.015 ± 0.00058 | 0.033 ± 0.0021 | 0.035 ± 0.0021 | 0.037 ± 0.0025 | 0.038 ± 0.0024 | 0.035 ± 0.0022 | 0.031 ± 0.0021 | 0.035 ± 0.0023 | 0.024 ± 0.0021 | 0.033 ± 0.0021 | 0.033 ± 0.0022 | 0.034 ± 0.0023 | 0.033 ± 0.0021 | 0.032 ± 0.002 | 0.03 ± 0.002 | 0.035 ± 0.0023 | 0.041 ± 0.0028 | 0.033 ± 0.0022 | 0.039 ± 0.0024 |
| Gen. from 6 DWIs fit tensor (subset 3) | 0.015 ± 0.00088 | 0.034 ± 0.002 | 0.035 ± 0.0022 | 0.036 ± 0.0025 | 0.037 ± 0.0025 | 0.031 ± 0.002 | 0.034 ± 0.0024 | 0.035 ± 0.0024 | 0.032 ± 0.002 | 0.035 ± 0.0024 | 0.037 ± 0.0028 | 0.034 ± 0.0023 | 0.037 ± 0.0023 | 0.035 ± 0.0024 | 0.034 ± 0.0023 | 0.033 ± 0.0021 | 0.034 ± 0.0022 | 0.033 ± 0.0022 | 0.034 ± 0.0021 |
| Gen. from 18 DWIs fit tensor | 0.011 ± 0.00087 | 0.017 ± 0.0011 | 0.019 ± 0.0012 | 0.018 ± 0.0012 | 0.019 ± 0.0013 | 0.018 ± 0.0012 | 0.017 ± 0.0012 | 0.018 ± 0.0012 | 0.019 ± 0.0013 | 0.019 ± 0.0013 | 0.018 ± 0.0012 | 0.018 ± 0.0012 | 0.019 ± 0.0012 | 0.018 ± 0.0012 | 0.016 ± 0.0011 | 0.018 ± 0.0011 | 0.021 ± 0.0015 | 0.018 ± 0.0012 | 0.02 ± 0.0013 |
| Supervised (w/ ext. clean data) | 0.0099 ± 0.00073 | 0.012 ± 0.00072 | 0.012 ± 0.00072 | 0.012 ± 0.00073 | 0.012 ± 0.00074 | 0.012 ± 0.00064 | 0.012 ± 0.00079 | 0.012 ± 0.00077 | 0.012 ± 0.00073 | 0.013 ± 0.00064 | 0.012 ± 0.00074 | 0.013 ± 0.00091 | 0.012 ± 0.00077 | 0.013 ± 0.00067 | 0.012 ± 0.0008 | 0.012 ± 0.00067 | 0.013 ± 0.0008 | 0.012 ± 0.00084 | 0.013 ± 0.00077 |
| SDnDTI (subset 1) | 0.014 ± 0.00091 | 0.018 ± 0.0011 | 0.018 ± 0.0011 | 0.018 ± 0.0011 | 0.018 ± 0.0012 | 0.018 ± 0.001 | 0.018 ± 0.0011 | 0.018 ± 0.0011 | 0.019 ± 0.0013 | 0.019 ± 0.0011 | 0.018 ± 0.0011 | 0.018 ± 0.0011 | 0.018 ± 0.0011 | 0.018 ± 0.0011 | 0.017 ± 0.001 | 0.018 ± 0.00098 | 0.021 ± 0.0014 | 0.018 ± 0.0011 | 0.02 ± 0.0013 |
| SDnDTI (subset 2) | 0.013 ± 0.00087 | 0.018 ± 0.0011 | 0.018 ± 0.00099 | 0.019 ± 0.0012 | 0.02 ± 0.0012 | 0.018 ± 0.0011 | 0.017 ± 0.0011 | 0.018 ± 0.0012 | 0.018 ± 0.0011 | 0.018 ± 0.001 | 0.017 ± 0.001 | 0.018 ± 0.0011 | 0.018 ± 0.001 | 0.018 ± 0.001 | 0.017 ± 0.001 | 0.019 ± 0.0012 | 0.02 ± 0.0014 | 0.018 ± 0.0011 | 0.019 ± 0.001 |
| SDnDTI (subset 3) | 0.013 ± 0.00078 | 0.018 ± 0.00094 | 0.018 ± 0.0011 | 0.018 ± 0.001 | 0.018 ± 0.0012 | 0.018 ± 0.00097 | 0.018 ± 0.0011 | 0.018 ± 0.001 | 0.018 ± 0.001 | 0.018 ± 0.001 | 0.019 ± 0.0013 | 0.018 ± 0.0011 | 0.019 ± 0.0011 | 0.018 ± 0.001 | 0.018 ± 0.0011 | 0.017 ± 0.00095 | 0.018 ± 0.001 | 0.018 ± 0.0011 | 0.018 ± 0.001 |
| SDnDTI (3 subsets average) | 0.01 ± 0.00082 | 0.012 ± 0.00065 | 0.012 ± 0.00073 | 0.012 ± 0.00065 | 0.014 ± 0.00089 | 0.012 ± 0.00071 | 0.012 ± 0.0007 | 0.012 ± 0.0007 | 0.013 ± 0.00089 | 0.013 ± 0.00067 | 0.013 ± 0.0007 | 0.012 ± 0.00068 | 0.012 ± 0.00067 | 0.013 ± 0.00075 | 0.012 ± 0.00064 | 0.012 ± 0.00067 | 0.014 ± 0.00088 | 0.012 ± 0.00076 | 0.013 ± 0.00074 |
| BM4D | 0.014 ± 0.00063 | 0.023 ± 0.0013 | 0.023 ± 0.0012 | 0.023 ± 0.0013 | 0.024 ± 0.0013 | 0.024 ± 0.0014 | 0.023 ± 0.0014 | 0.024 ± 0.0015 | 0.024 ± 0.0013 | 0.023 ± 0.0013 | 0.023 ± 0.0014 | 0.024 ± 0.0015 | 0.024 ± 0.0012 | 0.024 ± 0.0015 | 0.024 ± 0.0014 | 0.023 ± 0.0013 | 0.024 ± 0.0014 | 0.024 ± 0.0014 | 0.023 ± 0.0013 |
| AONLM | 0.01 ± 0.0007 | 0.023 ± 0.0012 | 0.024 ± 0.0012 | 0.024 ± 0.0012 | 0.024 ± 0.0013 | 0.024 ± 0.0013 | 0.024 ± 0.0013 | 0.025 ± 0.0014 | 0.024 ± 0.0013 | 0.023 ± 0.0013 | 0.024 ± 0.0013 | 0.025 ± 0.0014 | 0.024 ± 0.0012 | 0.024 ± 0.0015 | 0.025 ± 0.0014 | 0.024 ± 0.0012 | 0.024 ± 0.0013 | 0.024 ± 0.0012 | 0.024 ± 0.0012 |

**Supplementary Table S1. Mean absolute error.** The group mean (± the group standard deviation) of the mean absolute error between denoised images from different methods and the ground-truth images across the 20 HCP subjects for the b = 0 image volume and 18 diffusion-weighted image volumes.

| PSNR | b = 0 | DWI 1 | DWI 2 | DWI 3 | DWI 4 | DWI 5 | DWI 6 | DWI 7 | DWI 8 | DWI 9 | DWI 10 | DWI 11 | DWI 12 | DWI 13 | DWI 14 | DWI 15 | DWI 16 | DWI 17 | DWI 18 |
|---|---|---|---|---|---|---|---|---|---|---|---|---|---|---|---|---|---|---|---|
| Raw | 33.41 ± 0.6 | 27.26 ± 0.6 | 27.13 ± 0.57 | 27.20 ± 0.59 | 27.16 ± 0.59 | 27.27 ± 0.57 | 27.34 ± 0.59 | 26.80 ± 0.6 | 27.09 ± 0.56 | 27.34 ± 0.56 | 27.34 ± 0.58 | 27.07 ± 0.6 | 27.23 ± 0.55 | 26.79 ± 0.62 | 27.07 ± 0.61 | 27.34 ± 0.55 | 27.15 ± 0.57 | 27.22 ± 0.6 | 27.17 ± 0.55 |
| Gen. from 6 DWIs fit tensor (subset 1) | 32.19 ± 0.66 | 26.98 ± 0.66 | 26.89 ± 0.61 | 26.92 ± 0.63 | 26.92 ± 0.62 | 26.97 ± 0.6 | 27.04 ± 0.63 | 26.88 ± 0.62 | 26.29 ± 0.63 | 25.46 ± 0.77 | 26.82 ± 0.7 | 26.74 ± 0.63 | 26.49 ± 0.65 | 27.08 ± 0.65 | 27.46 ± 0.71 | 26.89 ± 0.59 | 24.92 ± 0.63 | 27.01 ± 0.63 | 25.40 ± 0.65 |
| Gen. from 6 DWIs fit tensor (subset 2) | 32.94 ± 0.58 | 26.99 ± 0.61 | 26.34 ± 0.61 | 25.96 ± 0.69 | 25.72 ± 0.63 | 26.51 ± 0.62 | 27.51 ± 0.67 | 26.52 ± 0.68 | 26.81 ± 0.63 | 27.10 ± 0.63 | 27.08 ± 0.66 | 26.77 ± 0.62 | 26.97 ± 0.59 | 27.13 ± 0.63 | 27.80 ± 0.68 | 26.31 ± 0.61 | 25.11 ± 0.69 | 26.90 ± 0.63 | 25.43 ± 0.62 |
| Gen. from 6 DWIs fit tensor (subset 3) | 33.41 ± 0.6 | 26.75 ± 0.62 | 26.45 ± 0.71 | 26.02 ± 0.67 | 25.96 ± 0.66 | 27.43 ± 0.65 | 26.55 ± 0.69 | 26.47 ± 0.66 | 27.19 ± 0.67 | 26.47 ± 0.67 | 25.95 ± 0.71 | 26.73 ± 0.64 | 25.90 ± 0.63 | 26.58 ± 0.66 | 26.79 ± 0.56 | 27.05 ± 0.62 | 26.93 ± 0.63 | 26.98 ± 0.65 | 26.86 ± 0.63 |
| Gen. from 18 DWIs fit tensor | 35.31 ± 0.75 | 32.07 ± 0.92 | 31.55 ± 0.8 | 31.73 ± 0.83 | 31.32 ± 0.83 | 31.87 ± 0.87 | 32.17 ± 0.95 | 32.05 ± 0.86 | 31.57 ± 0.8 | 31.39 ± 0.83 | 31.98 ± 0.98 | 31.88 ± 0.88 | 31.50 ± 0.84 | 32.02 ± 0.88 | 32.61 ± 1.04 | 31.87 ± 0.87 | 30.84 ± 0.74 | 31.91 ± 0.9 | 30.91 ± 0.75 |
| Supervised (w/ ext. clean data) | 36.54 ± 0.76 | 36.08 ± 0.56 | 35.83 ± 0.54 | 35.90 ± 0.53 | 35.71 ± 0.53 | 35.99 ± 0.5 | 35.91 ± 0.6 | 36.00 ± 0.56 | 35.85 ± 0.54 | 35.58 ± 0.49 | 36.09 ± 0.57 | 35.41 ± 0.6 | 35.82 ± 0.57 | 35.55 ± 0.47 | 36.20 ± 0.61 | 35.91 ± 0.51 | 35.06 ± 0.52 | 35.74 ± 0.61 | 35.55 ± 0.56 |
| SDnDTI (subset 1) | 33.18 ± 0.69 | 31.90 ± 0.97 | 31.87 ± 0.86 | 31.95 ± 0.94 | 31.73 ± 0.81 | 31.84 ± 0.87 | 31.94 ± 0.95 | 31.76 ± 0.96 | 31.31 ± 0.81 | 31.04 ± 1.22 | 31.77 ± 1.06 | 31.81 ± 0.9 | 31.62 ± 0.98 | 31.97 ± 0.87 | 32.14 ± 1.11 | 31.79 ± 0.87 | 30.54 ± 0.89 | 31.75 ± 0.92 | 31.10 ± 0.85 |
| SDnDTI (subset 2) | 34.00 ± 0.6 | 31.99 ± 0.91 | 31.68 ± 0.93 | 31.17 ± 1.06 | 31.14 ± 0.93 | 31.50 ± 0.96 | 32.13 ± 1.02 | 31.56 ± 1.04 | 31.68 ± 0.97 | 31.95 ± 0.88 | 32.13 ± 1.02 | 31.74 ± 0.93 | 31.96 ± 0.88 | 32.02 ± 0.93 | 32.49 ± 1.04 | 31.28 ± 0.95 | 30.76 ± 1.01 | 31.74 ± 0.94 | 31.12 ± 0.91 |
| SDnDTI (subset 3) | 34.35 ± 0.63 | 31.71 ± 1.01 | 31.70 ± 1.09 | 31.63 ± 0.96 | 31.18 ± 0.84 | 32.16 ± 0.92 | 31.94 ± 1.06 | 31.91 ± 0.9 | 31.91 ± 0.96 | 31.63 ± 0.92 | 31.75 ± 0.96 | 31.75 ± 1.01 | 31.24 ± 1.06 | 31.79 ± 0.91 | 31.80 ± 0.99 | 32.05 ± 0.94 | 31.75 ± 0.81 | 31.88 ± 0.93 | 31.82 ± 1.01 |
| SDnDTI (3 subsets average) | 35.85 ± 0.77 | 34.71 ± 1.43 | 34.50 ± 1.32 | 34.42 ± 1.41 | 33.82 ± 1.20 | 34.41 ± 1.34 | 34.61 ± 1.48 | 34.54 ± 1.37 | 34.09 ± 1.22 | 34.20 ± 1.38 | 34.42 ± 1.47 | 34.48 ± 1.43 | 34.50 ± 1.44 | 34.49 ± 1.28 | 34.86 ± 1.56 | 34.33 ± 1.36 | 33.70 ± 1.19 | 34.54 ± 1.39 | 34.13 ± 1.30 |
| BM4D | 33.67 ± 0.46 | 30.35 ± 0.51 | 30.23 ± 0.47 | 30.25 ± 0.46 | 30.19 ± 0.49 | 30.36 ± 0.5 | 30.36 ± 0.53 | 29.96 ± 0.58 | 30.10 ± 0.46 | 30.35 ± 0.49 | 30.33 ± 0.51 | 29.94 ± 0.56 | 30.17 ± 0.46 | 29.95 ± 0.59 | 29.94 ± 0.53 | 30.36 ± 0.49 | 30.18 ± 0.5 | 30.16 ± 0.53 | 30.25 ± 0.47 |
| AONLM | 34.12 ± 0.55 | 30.24 ± 0.49 | 30.16 ± 0.46 | 30.13 ± 0.45 | 30.15 ± 0.46 | 30.08 ± 0.47 | 30.24 ± 0.51 | 29.80 ± 0.55 | 30.04 ± 0.47 | 30.27 ± 0.47 | 30.23 ± 0.49 | 29.82 ± 0.54 | 30.09 ± 0.44 | 29.88 ± 0.55 | 29.82 ± 0.51 | 30.24 ± 0.47 | 30.11 ± 0.48 | 30.06 ± 0.5 | 30.17 ± 0.46 |

**Supplementary Table S2. Peak signal-to-noise ratio.** The group mean (± the group standard deviation) of the peak signal-to-noise ratio (PSNR) between denoised images from different methods and the ground-truth images across the 20 HCP subjects for the b = 0 image volume and 18 diffusion-weighted image volumes.

| SSIM | b = 0 | DWI 1 | DWI 2 | DWI 3 | DWI 4 | DWI 5 | DWI 6 | DWI 7 | DWI 8 | DWI 9 | DWI 10 | DWI 11 | DWI 12 | DWI 13 | DWI 14 | DWI 15 | DWI 16 | DWI 17 | DWI 18 |
|---|---|---|---|---|---|---|---|---|---|---|---|---|---|---|---|---|---|---|---|
| Raw | 0.97 ± 0.0039 | 0.89 ± 0.012 | 0.88 ± 0.013 | 0.89 ± 0.012 | 0.87 ± 0.013 | 0.89 ± 0.011 | 0.9 ± 0.011 | 0.88 ± 0.012 | 0.87 ± 0.013 | 0.89 ± 0.012 | 0.89 ± 0.012 | 0.89 ± 0.012 | 0.89 ± 0.012 | 0.87 ± 0.014 | 0.89 ± 0.012 | 0.89 ± 0.011 | 0.88 ± 0.012 | 0.89 ± 0.012 | 0.89 ± 0.012 |
| Gen. from 6 DWIs fit tensor (subset 1) | 0.96 ± 0.0042 | 0.89 ± 0.012 | 0.88 ± 0.013 | 0.89 ± 0.012 | 0.87 ± 0.014 | 0.89 ± 0.012 | 0.89 ± 0.012 | 0.88 ± 0.012 | 0.86 ± 0.016 | 0.85 ± 0.015 | 0.88 ± 0.012 | 0.89 ± 0.012 | 0.88 ± 0.013 | 0.87 ± 0.014 | 0.9 ± 0.011 | 0.89 ± 0.012 | 0.83 ± 0.016 | 0.89 ± 0.011 | 0.85 ± 0.016 |
| Gen. from 6 DWIs fit tensor (subset 2) | 0.97 ± 0.004 | 0.89 ± 0.012 | 0.86 ± 0.013 | 0.87 ± 0.013 | 0.84 ± 0.015 | 0.88 ± 0.012 | 0.91 ± 0.011 | 0.88 ± 0.013 | 0.87 ± 0.013 | 0.88 ± 0.012 | 0.89 ± 0.012 | 0.89 ± 0.012 | 0.89 ± 0.012 | 0.87 ± 0.013 | 0.91 ± 0.01 | 0.88 ± 0.013 | 0.84 ± 0.016 | 0.89 ± 0.012 | 0.85 ± 0.015 |
| Gen. from 6 DWIs fit tensor (subset 3) | 0.97 ± 0.0039 | 0.89 ± 0.012 | 0.87 ± 0.013 | 0.87 ± 0.013 | 0.85 ± 0.015 | 0.89 ± 0.011 | 0.89 ± 0.013 | 0.87 ± 0.013 | 0.88 ± 0.013 | 0.87 ± 0.014 | 0.87 ± 0.014 | 0.89 ± 0.012 | 0.87 ± 0.013 | 0.86 ± 0.014 | 0.89 ± 0.012 | 0.89 ± 0.011 | 0.88 ± 0.012 | 0.89 ± 0.012 | 0.88 ± 0.012 |
| Gen. from 18 DWIs fit tensor | 0.99 ± 0.0023 | 0.96 ± 0.0045 | 0.95 ± 0.0054 | 0.96 ± 0.005 | 0.95 ± 0.0059 | 0.96 ± 0.0045 | 0.96 ± 0.0043 | 0.96 ± 0.0047 | 0.95 ± 0.0059 | 0.95 ± 0.0056 | 0.96 ± 0.0047 | 0.96 ± 0.0047 | 0.96 ± 0.0049 | 0.95 ± 0.0055 | 0.97 ± 0.004 | 0.96 ± 0.0046 | 0.95 ± 0.0063 | 0.96 ± 0.0045 | 0.95 ± 0.0059 |
| Supervised (w/ ext. clean data) | 0.99 ± 0.0019 | 0.98 ± 0.0019 | 0.98 ± 0.0023 | 0.98 ± 0.0021 | 0.98 ± 0.0024 | 0.98 ± 0.0019 | 0.98 ± 0.0019 | 0.98 ± 0.0021 | 0.98 ± 0.0026 | 0.98 ± 0.0022 | 0.98 ± 0.002 | 0.98 ± 0.002 | 0.98 ± 0.002 | 0.98 ± 0.0024 | 0.98 ± 0.0018 | 0.98 ± 0.002 | 0.98 ± 0.0024 | 0.98 ± 0.0019 | 0.98 ± 0.0024 |
| SDnDTI (subset 1) | 0.98 ± 0.0029 | 0.96 ± 0.0041 | 0.96 ± 0.0047 | 0.96 ± 0.0043 | 0.96 ± 0.0048 | 0.96 ± 0.0039 | 0.96 ± 0.0041 | 0.96 ± 0.0041 | 0.95 ± 0.0057 | 0.95 ± 0.0049 | 0.96 ± 0.0042 | 0.96 ± 0.0042 | 0.96 ± 0.0043 | 0.96 ± 0.0049 | 0.97 ± 0.0037 | 0.96 ± 0.0041 | 0.95 ± 0.0054 | 0.96 ± 0.004 | 0.95 ± 0.0052 |
| SDnDTI (subset 2) | 0.98 ± 0.0025 | 0.96 ± 0.004 | 0.96 ± 0.0047 | 0.95 ± 0.0047 | 0.95 ± 0.0052 | 0.96 ± 0.0042 | 0.97 ± 0.0037 | 0.96 ± 0.0045 | 0.96 ± 0.005 | 0.96 ± 0.0042 | 0.96 ± 0.004 | 0.96 ± 0.0041 | 0.96 ± 0.0038 | 0.96 ± 0.0046 | 0.97 ± 0.0035 | 0.96 ± 0.0044 | 0.95 ± 0.0054 | 0.96 ± 0.0042 | 0.95 ± 0.0049 |
| SDnDTI (subset 3) | 0.98 ± 0.0024 | 0.96 ± 0.0039 | 0.96 ± 0.0047 | 0.96 ± 0.0043 | 0.96 ± 0.0053 | 0.96 ± 0.0038 | 0.96 ± 0.0041 | 0.96 ± 0.004 | 0.96 ± 0.0049 | 0.96 ± 0.0046 | 0.96 ± 0.0044 | 0.96 ± 0.0041 | 0.96 ± 0.0042 | 0.95 ± 0.005 | 0.96 ± 0.0039 | 0.96 ± 0.0038 | 0.96 ± 0.0044 | 0.96 ± 0.004 | 0.96 ± 0.0043 |
| SDnDTI (3 subsets average) | 0.98 ± 0.0018 | 0.98 ± 0.0019 | 0.98 ± 0.0024 | 0.98 ± 0.0021 | 0.97 ± 0.0026 | 0.98 ± 0.002 | 0.98 ± 0.002 | 0.98 ± 0.002 | 0.98 ± 0.0027 | 0.98 ± 0.0023 | 0.98 ± 0.0021 | 0.98 ± 0.0021 | 0.98 ± 0.002 | 0.98 ± 0.0024 | 0.98 ± 0.0018 | 0.98 ± 0.0021 | 0.98 ± 0.0025 | 0.98 ± 0.002 | 0.98 ± 0.0024 |
| BM4D | 0.94 ± 0.002 | 0.94 ± 0.0056 | 0.94 ± 0.0056 | 0.94 ± 0.0054 | 0.93 ± 0.0063 | 0.94 ± 0.0057 | 0.94 ± 0.0056 | 0.94 ± 0.006 | 0.93 ± 0.0064 | 0.94 ± 0.0059 | 0.94 ± 0.0055 | 0.94 ± 0.0062 | 0.94 ± 0.0054 | 0.93 ± 0.0074 | 0.94 ± 0.0056 | 0.94 ± 0.0053 | 0.94 ± 0.006 | 0.94 ± 0.0058 | 0.94 ± 0.0057 |
| AONLM | 0.98 ± 0.0022 | 0.94 ± 0.0056 | 0.94 ± 0.0058 | 0.94 ± 0.0055 | 0.93 ± 0.0064 | 0.94 ± 0.0053 | 0.94 ± 0.0056 | 0.94 ± 0.0061 | 0.93 ± 0.0066 | 0.94 ± 0.0058 | 0.94 ± 0.0055 | 0.94 ± 0.0054 | 0.94 ± 0.0054 | 0.93 ± 0.0073 | 0.94 ± 0.0053 | 0.94 ± 0.0053 | 0.93 ± 0.0059 | 0.94 ± 0.0057 | 0.94 ± 0.0057 |

**Supplementary Table S3. Structural similarity index.** The group mean (± the group standard deviation) of the structural similarity index (SSIM) between denoised images from different methods and the ground-truth images across the 20 HCP subjects for the b = 0 image volume and 18 diffusion-weighted image volumes.

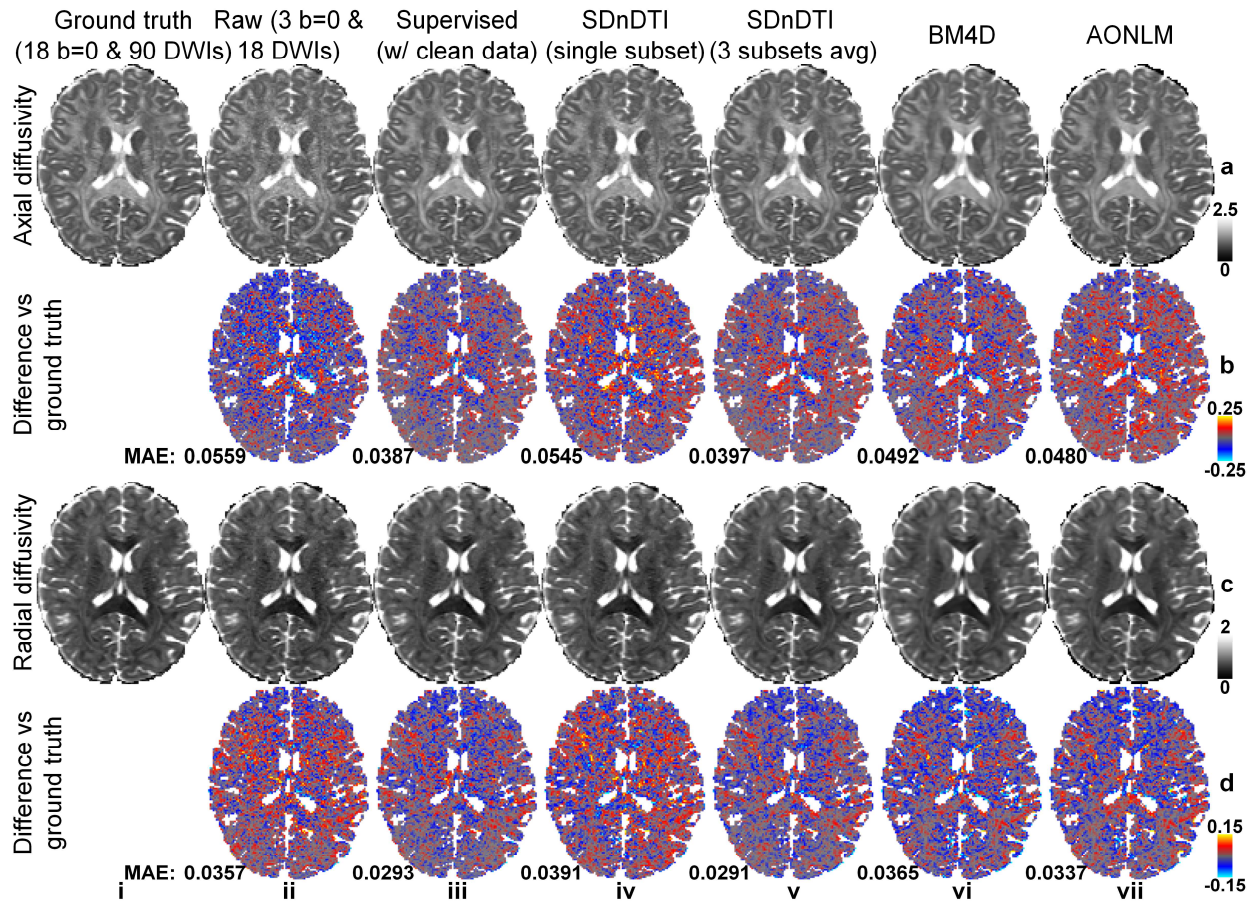

**Supplementary Figure S3. Axial and radial diffusivity.** Maps of axial diffusivity (row a) and radial diffusivity (row c) derived from the diffusion tensors fitted using all 18 b = 0 and 90 diffusion-weighted images (DWIs) (ground truth, column i), raw data consisting of three b = 0 and 18 DWIs (column ii), the raw data denoised by supervised learning with the ground-truth images as the training target (i.e., supervised denoising) (column iii), the subset 2 of SDnDTI-denoised data (column iv), the average of all three subsets of SDnDTI-denoised data (column v), and the raw data denoised by BM4D (column vi) and AONLM (column vi), and their residual maps (rows b, d) compared to the ground-truth maps from a representative HCP subject. The mean absolute difference (MAD) of each map compared to the ground truth within the brain (excluding the cerebrospinal fluid) is displayed at the bottom of the residual map. The unit of the diffusivity is μm²/ms.

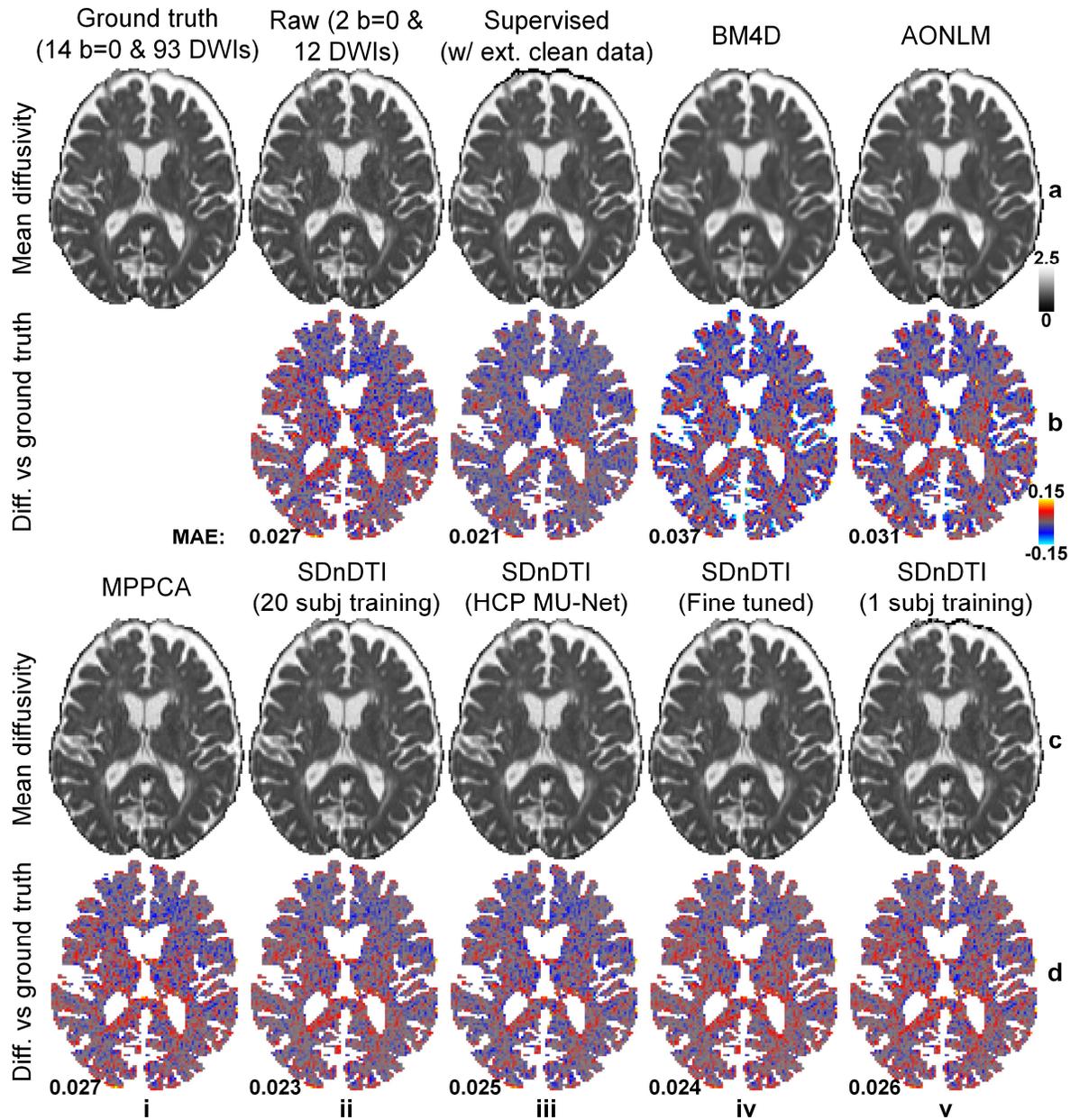

**Supplementary Figure S4. Mean diffusivity.** Maps of mean diffusivity (row a, c) derived from the diffusion tensors fitted using all 14 b = 0 and 93 diffusion-weighted images (DWIs) (ground truth, a, i), raw data consisting of two b = 0 and 12 DWIs (a, ii), the raw data denoised by supervised learning with the ground-truth DWIs as the training target (i.e., supervised denoising) (a, iii), BM4D (a, iv), AONLM (a, v) and MPPCA (c, i), and SDnDTI (c, ii–v), and their residual maps (rows b, d) compared to the ground-truth map from a representative HCP-A subject. SDnDTI results were generated by an MU-Net trained on the data from 20 HCP-A subjects (c, ii), an MU-Net trained on the data from 20 HCP subjects (c, iii), an MU-Net with parameters from the MU-Net trained on the data from 20 HCP subjects as initialization and further fine-tuned using the data of each HCP-A subject (c, iv), and an MU-Net trained on the data from the data of each HCP-A subject (c, v). The mean absolute difference (MAD) of each map compared to the ground truth within the brain (excluding the cerebrospinal fluid) is displayed at the bottom of the residual map. The unit of the diffusivity is μm²/ms.

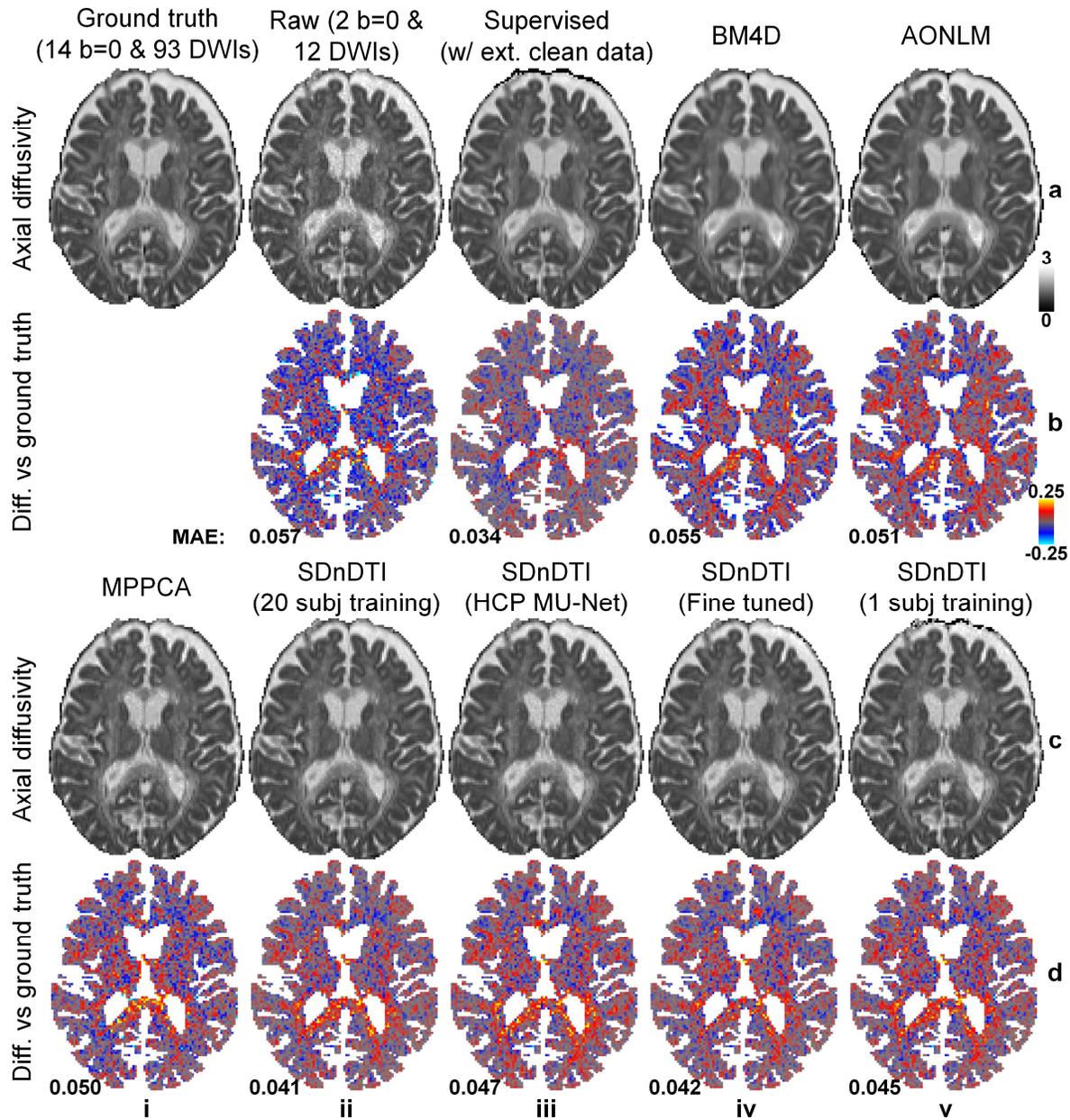

**Supplementary Figure S5. Axial diffusivity.** Maps of axial diffusivity (row a, c) derived from the diffusion tensors fitted using all 14 b = 0 and 93 diffusion-weighted images (DWIs) (ground truth, a, i), raw data consisting of two b = 0 and 12 DWIs (a, ii), the raw data denoised by supervised learning with the ground-truth DWIs as the training target (i.e., supervised denoising) (a, iii), BM4D (a, iv), AONLM (a, v) and MPPCA (c, i), and SDnDTI (c, ii–v), and their residual maps (rows b, d) compared to the ground-truth map from a representative HCP-A subject. SDnDTI results were generated by an MU-Net trained on the data from 20 HCP-A subjects (c, ii), an MU-Net trained on the data from 20 HCP subjects (c, iii), an MU-Net with parameters from the MU-Net trained on the data from 20 HCP subjects as initialization and further fine-tuned using the data of each HCP-A subject (c, iv), and an MU-Net trained on the data from the data of each HCP-A subject (c, v). The mean absolute difference (MAD) of each map compared to the ground truth within the brain (excluding the cerebrospinal fluid) is displayed at the bottom of the residual map. The unit of the diffusivity is µm²/ms.

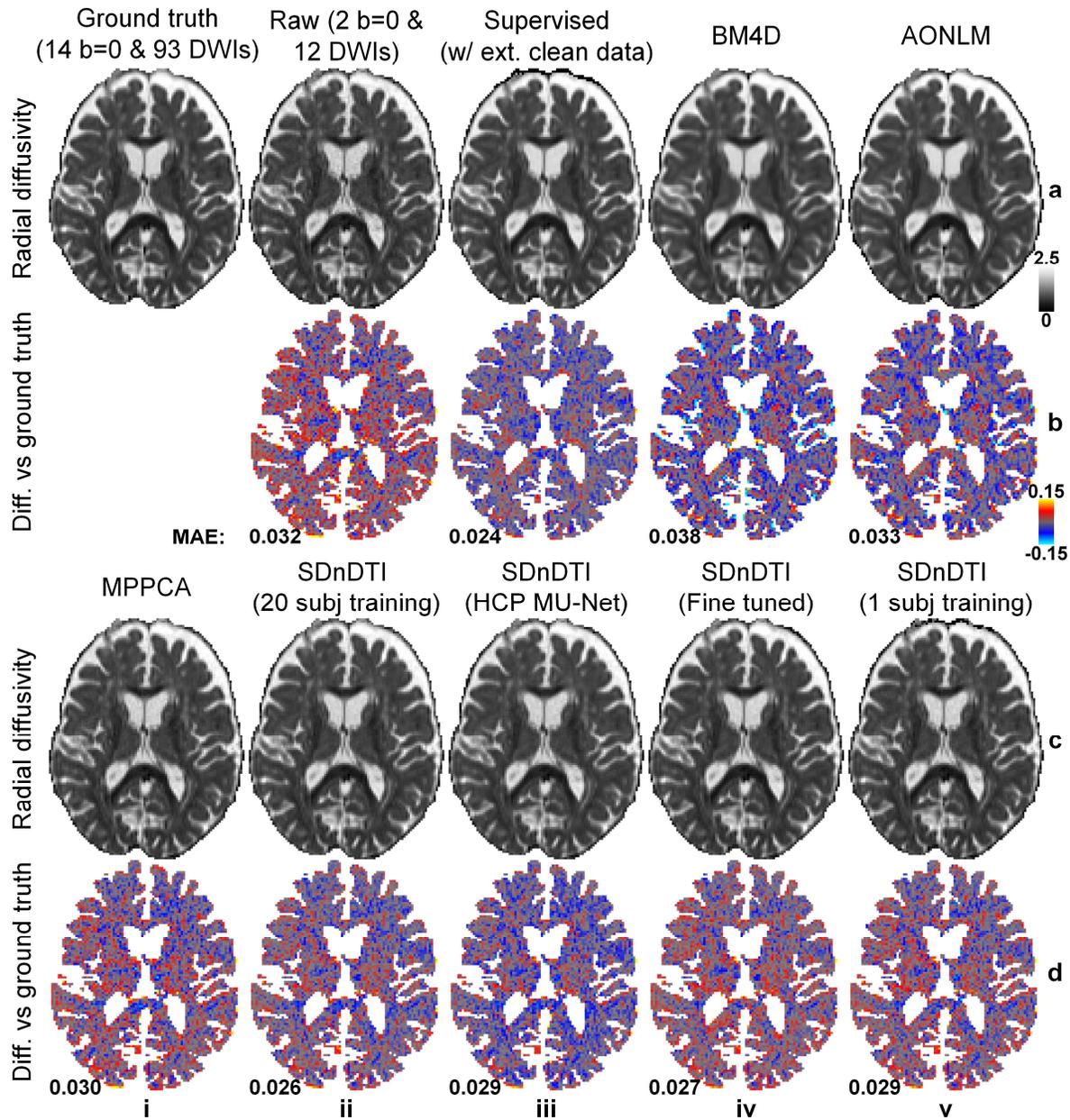

**Supplementary Figure S6. Radial diffusivity.** Maps of axial diffusivity (row a, c) derived from the diffusion tensors fitted using all 14 b = 0 and 93 diffusion-weighted images (DWIs) (ground truth, a, i), raw data consisting of two b = 0 and 12 DWIs (a, ii), the raw data denoised by supervised learning with the ground-truth DWIs as the training target (i.e., supervised denoising) (a, iii), BM4D (a, iv), AONLM (a, v) and MPPCA (c, i), and SDnDTI (c, ii–v), and their residual maps (rows b, d) compared to the ground-truth map from a representative HCP-A subject. SDnDTI results were generated by an MU-Net trained on the data from 20 HCP-A subjects (c, ii), an MU-Net trained on the data from 20 HCP subjects (c, iii), an MU-Net with parameters from the MU-Net trained on the data from 20 HCP subjects as initialization and further fine-tuned using the data of each HCP-A subject (c, iv), and an MU-Net trained on the data from the data of each HCP-A subject (c, v). The mean absolute difference (MAD) of each map compared to the ground truth within the brain (excluding the cerebrospinal fluid) is displayed at the bottom of the residual map. The unit of the diffusivity is μm²/ms.